\documentclass[acmsmall,screen]{acmart}

\usepackage{amsmath,amssymb,amsfonts}
\usepackage{algorithmic}
\usepackage{graphicx}
\usepackage{textcomp}
\usepackage{makecell}
\usepackage{xcolor}
\usepackage{booktabs}
\usepackage{tikz}
\usepackage{ulem}
\usepackage{multirow}
\usepackage{pifont}
\usepackage{tcolorbox}
\usepackage{soul}
\usepackage{setspace}
\usepackage{threeparttable}
\usepackage{url}
\usepackage{subcaption}
\usepackage{graphicx}

\newcommand{\finding}[2]{
    \begin{center}
    \fcolorbox{black}{gray!10}{\parbox{.97\linewidth}{
    {#2}
    }}
    \end{center}
}

\AtBeginDocument{%
  }

\setcopyright{acmlicensed}
\copyrightyear{XXXXXXX}
\acmYear{XXXXXXX}
\acmDOI{XXXXXXX.XXXXXXX}

\makeatletter
\def\@journalName{} 
\def\@journalYear{} 
\def\@acmArticle{}  
\def\@acmVolume{}   
\def\@acmNumber{}   
\def\@acmArticleSeq{} 
\def\@acmPubDate{}  
\makeatother

\begin{document}

\title{Deep Learning Framework Testing via Model Mutation: How Far Are We?}

\author{Yanzhou Mu}
\authornote{Both authors contributed equally to this research.}
\email{602022320006@smail.nju.edu.cn}
\affiliation{
  \institution{State Key Laboratory for Novel Software Technology, Nanjing University}
  \city{Nanjing}
  \state{Jiangsu}
  \country{China}
}

\author{Rong Wang}
\authornotemark[1]
\email{wangrong_hcir@163.com}
\affiliation{
  \institution{School of Information Science and Technology, Nantong University}
  \city{Nantong}
  \state{Jiangsu}
  \country{China}
}

\author{Juan Zhai}
\email{juanzhai@umass.edu}
\affiliation{
  \institution{University of Massachusetts Amherst}
  \city{Amherst}
  \state{Massachusetts}
  \country{USA}
}

\author{Chunrong Fang}
\authornote{Corresponding author.}
\email{fangchunrong@nju.edu.cn}
\affiliation{
  \institution{State Key Laboratory for Novel Software Technology, Nanjing University}
  \city{Nanjing}
  \state{Jiangsu}
  \country{China}
}

\author{Xiang Chen}
\email{xchencs@ntu.edu.cn}
\affiliation{
  \institution{School of Artificial Intelligence and Computer Science, Nantong University}
  \city{Nantong}
  \state{Jiangsu}
  \country{China}
}

\author{Zhiyuan Peng}
\email{522022000027@smail.nju.edu.cn}
\affiliation{
  \institution{State Key Laboratory for Novel Software Technology, Nanjing University}
  \city{Nanjing}
  \state{Jiangsu}
  \country{China}
}

\author{Peiran Yang}
\email{522023320183@smail.nju.edu.cn}
\affiliation{
  \institution{State Key Laboratory for Novel Software Technology, Nanjing University}
  \city{Nanjing}
  \state{Jiangsu}
  \country{China}
}

\author{Ruixiang Qian}
\email{qianrx@smail.nju.edu.cn}
\affiliation{
  \institution{State Key Laboratory for Novel Software Technology, Nanjing University}
  \city{Nanjing}
  \state{Jiangsu}
  \country{China}
}

\author{Shaoyu Yang}
\email{shaoyuyoung@gmail.com}
\affiliation{
  \institution{State Key Laboratory for Novel Software Technology, Nanjing University}
  \city{Nanjing}
  \state{Jiangsu}
  \country{China}
}

\author{Zhenyu Chen}
\email{zychen@nju.edu.cn}
\affiliation{
  \institution{State Key Laboratory for Novel Software Technology, Nanjing University}
  \city{Nanjing}
  \state{Jiangsu}
  \country{China}
}

\renewcommand{\shortauthors}{MU et al.}

\begin{abstract}
Deep Learning (DL) frameworks are a fundamental component of DL development.
Therefore, the detection of DL framework defects is important and challenging.
As one of the most widely adopted DL testing techniques, model mutation has recently gained significant attention.
In this study, we revisit the defect detection ability of existing mutation-based testing methods and investigate the factors that influence their effectiveness.
To begin with, we reviewed existing methods and observed that many of them mutate DL models (e.g., changing their parameters) without any customization, ignoring the unique challenges in framework testing.
Another issue with these methods is their limited effectiveness, characterized by a high rate of false positives caused by illegal mutations arising from the use of generic, non-customized mutation operators.
Moreover, we tracked the defects identified by these methods and discovered that most of them were ignored by developers.
Motivated by these observations, we investigate the effectiveness of existing mutation-based testing methods in detecting important defects that have been authenticated by framework developers.
We begin by collecting defect reports from three popular frameworks and classifying them based on framework developers' ratings to build a comprehensive dataset. We then perform an in-depth analysis to uncover valuable insights.
Based on our findings, we propose optimization strategies to address the shortcomings of existing approaches. Following these optimizations, we identified seven new defects, four of which were confirmed by developers as high-priority issues, with three resolved.
In summary, we identified 39 unique defects across just 23 models, of which 31 were confirmed by developers, and eight have been fixed.
\end{abstract}



\maketitle

\section{Introduction}
\label{sec:introduction}

Deep learning (DL) software systems gradually undertake indispensable tasks in many 
life-critical fields, such as medical diagnosis~\cite{obermeyer2016predicting}, 
autonomous driving~\cite{ChenSKX15}, speech recognition~\cite{0040158, LeCunBH15}, and software 
engineering~\cite{0003HLXZHGXDZ19,0003HL0HGXDZ19,li2019deepfl,ZhangXLQZDXYCLC19}. The quality issues of DL software systems can cause huge economic losses and threaten people's life safety~\cite{teslanews}. 
The development, deployment, and execution of DL models deeply rely on DL frameworks, and the defects of the DL framework will incur quality risks in DL models and lead to undesired consequences. 
Therefore, the quality assurance of DL frameworks has become a hot research topic in recent years~\cite{zou2023ramos,wang2022eagle,guo2020audee,pham2019cradle,luo2021graphfuzz,li2022mmos,park2023gradfuzz,Han2020WhatDP,Yan2021ExposingNB,Xie2021LeveragingDT,Deng2022FuzzingDL,Yang2023FuzzingAD,Wei2022FreeLF,Chen2022TowardUD,Zhang2020AnES,Makkouk2022AnES,Tambon2021SilentBI}.

\begin{figure}[hbp]
     \centering
     \footnotesize
      \includegraphics[width=.95\linewidth]{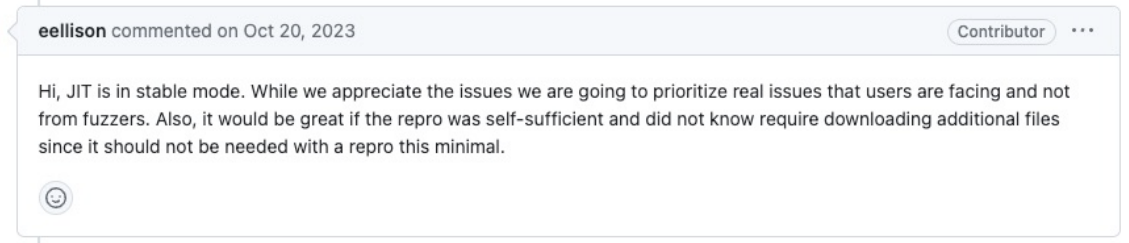}
     \caption{An example of developer response on defect report}
     \label{fig:demo}
\end{figure}

Some researchers ~\cite{pham2019cradle,wang2020lemon,guo2020audee,li2023comet} adopt
various mutation operators, such as modifying the model structure and editing layer parameters, on the DL models to generate model mutants and leverage the mutants as the test inputs for validating DL frameworks. 
To unveil potential defects, researchers analyze whether the generated mutants have inconsistent outputs, crashes, and NAN symptoms across different DL frameworks and report defects. In our study, such methods are called mutation-based testing methods.
Despite being effective in many cases, we found that \textbf{most of these methods use naive mutation operators}.
The test input of frameworks is DL models and their mutants. Promoting the diversity of test input to cover richer framework interfaces and more complex invocations of interfaces is critical for improving testing effectiveness. However, most existing methods ignore this characteristic during model mutation. 
Another downside is the \textbf{high false positives caused by illegal mutants.}
Existing methods lack sufficient constraints during mutation, resulting in many generated mutants crashing or the outputs exceeding the effective accuracy range, which is not caused by framework defects. Illegal mutants are generated models that crash or produce outputs beyond the effective accuracy range for the lack of insufficient constraints during mutation, which is not caused by framework defects. These are false positives and are rare in real-world scenarios, specifically in the deployment environments of DL models for industrial applications (e.g., face recognition), leading to huge manual inspection costs and limiting test efficiency.
Last but not least, 
\textbf{the defects detected by existing methods are often ignored by developers or rated as trivial.}
The developers find the defects reported by existing methods are far from the real user practice and consider the defects ``low priority''. Fig.~\ref{fig:demo} shows an example of a PyTorch developer's response to a defect submitted by existing methods. 
It indicated that the developers are more concerned about real-world defects.

Motivated by these observations, this paper aims to optimize existing mutation-based testing methods and further promote the quality of DL frameworks.
We achieve this by investigating the following three research questions: 

\noindent\(\bullet\)
\textbf{RQ1: What types of defects are developers concerned about?} 
Since developers prioritize important defects, this RQ aims to identify the types of framework defects that developers consider important.

To answer RQ1, we collect the issue reports from MindSpore, PyTorch, and TensorFlow up to the end of 2023, select 3,000 defect reports of ``High Priority'' based on the marked tags of developers, and note them as HP defects. 
Then, six volunteers label them and ultimately determine seven major types and 21 sub-types, achieving three key findings:
(1) the defects of managing resources like GPU, and memory are the most common sub-type of HP defects;
(2) compared to accuracy defects, which can be caused by randomness in training, developers are more concerned about performance defects, such as abnormal loss and evaluation metrics in model training and inference;
(3) crash defects are the most common type of HP defects, especially those about complex invocations of framework interfaces.

\noindent\(\bullet\)
\textbf{RQ2: How effective are the existing mutation-based testing methods in detecting defects?} 
This RQ aims to evaluate defect detection ability, especially the HP defects of existing mutation-based testing methods, based on the classification results of RQ1. We apply these methods on a more comprehensive benchmark with more DL models from various industry applications (e.g., object detection models in autonomous driving systems).

To answer RQ2, we apply existing mutation-based testing methods on three widely-used DL frameworks, i.e., TensorFlow, ONNX, and PyTorch. Then, we analyze the newly detected defects and the previously reported defects in their studies, especially analyzing the count and type of HP defects that existing methods can detect. 
Finally, we obtain three key findings: 
(1) existing methods rarely report inconsistency defects during execution while reporting many duplicate crash defects and NAN defects;
(2) existing methods can detect HP defects in five major types and eight sub-types, mainly including framework interface implementation defects and NAN defects;
(3) among missing sub-types of HP defects, the detection ability of existing methods for performance defects related to model training is the weakest.

\noindent\(\bullet\)
\textbf{RQ3: What factors affect the generated mutants of existing mutation-based testing methods?} 
This RQ aims to investigate the specific effects of the factors that affect the generated mutants since the mutants are the key to detecting defects. Furthermore, analyzing their impact factors can help to understand the causes of the shortcomings of existing methods. 

To answer RQ3, we identify three factors: mutation type, mutation order, and mutation position. Then we conduct the following experiments to investigate their specific effects on model mutation:
(1) compare the output inconsistency of the mutants generated by different types of mutation operators; 
(2) compare the output inconsistency of mutants generated by low-order and high-order mutation; 
(3) compare the output inconsistency of mutants generated by mutating middle layers from different parts of models. 
We make four key observations for this RQ: 
(1) the structure-mutation operators outperform others in model mutation;
(2) the input-mutation operators can detect defects related to efficiency and resource allocation;
(3) Low-order mutation outperforms high-order mutation and considering higher mutation orders tend to introduce more false positives;
(4) mutating the layers of the backbone of models outperforms mutating in the task head of models.

Based on the above analysis, we conclude three aspects of optimization suggestions: 
(1) design new mutation operators that are similar to the real practice of developers;
(2) enhance the mutation constraints to avoid generating invalid models;
(3) cover more defect types, such as performance defects and resource scheduling defects exposed in model training.
Our naive implementation of several improvements detects seven new defects, out of which, two are new types of defects, i.e., efficiency defects and memory management defects. 
This proves the practical value of our suggestions. We show detailed experimental results on our website~\cite{sharelink} for replication. 

Our key contributions are as follows:


\begin{itemize}
    \item \textbf{Comprehensive Benchmark.} We collect 11 DL models widely used in eight industry tasks and 12 from existing work. The 11 models can support the practical significance of this study, while the 12 most commonly used models from existing work ensure fairness.
    \item \textbf{Innovative Perspective.} We leverage developers' expertise to categorize defects into different groups and further evaluate the defect detection ability of existing mutation-based testing methods based on the defect classification.
    \item \textbf{Practical Optimizations.} 
    We conclude effective optimization strategies for mutation-based testing methods from three aspects: mutation innovation, constraint enhancement, and detection extension. 
    \item \textbf{New Findings}.
    We detected 39 new framework defects during our study, of which 31 have been confirmed, and eight have been fixed.
    Notably, seven defects were detected with the help of our optimization strategies, and 32 defects were detected with our newly constructed benchmark and profound analysis.
\end{itemize}

The organization of this paper is as follows. Section 1 introduces the background of DL framework testing, analyzes the shortcomings of existing work and then briefly presents how we conduct this study and our experimental findings. Section 2 introduces the basic knowledge of the model-level DL framework methods. Section 3 presents the entire details of our study, while Section 4 introduces the experimental settings, and Section 5 shows the results and findings, respectively. We further discuss the optimization suggestions of existing methods based on the findings and present the typical bug cases after optimization in Section 6. Section 7 introduces the threats to the validity of our work. Section 8 presents the related work of DL framework testing, and Section 9 briefly summarizes our work.

\section{Background}
\label{sec:background}

\subsection{Testing DL Frameworks via DL Models}

Researchers adopt DL models as test inputs for DL frameworks, with model generation as the core of testing. Existing methods are divided into (1) mutation-based testing and (2) template-based testing. Our research focuses on the first type.

\begin{figure}[]
     \centering
     \footnotesize
    \includegraphics[width=0.95\textwidth]{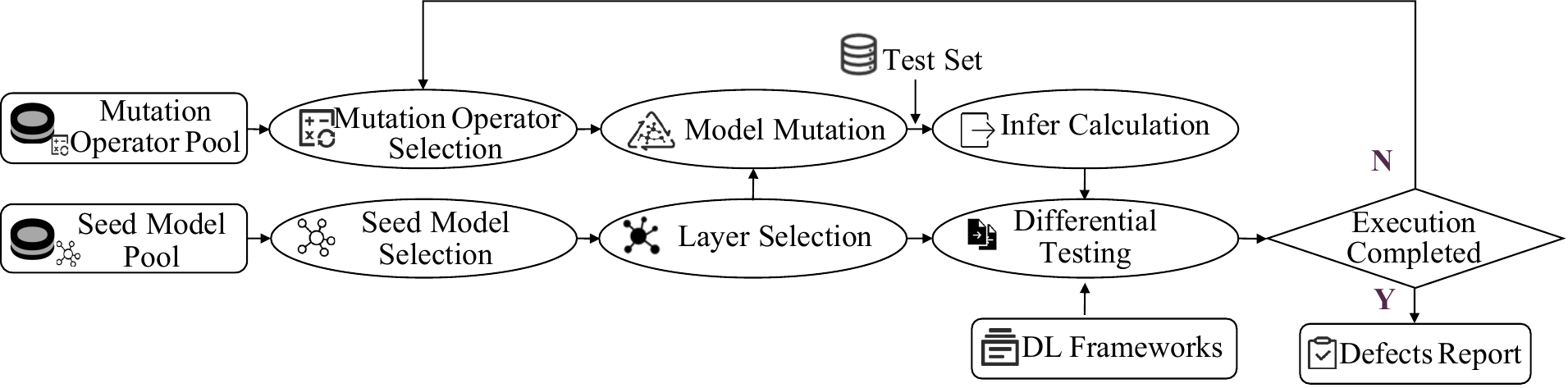}
     \caption{Workflow of mutation-based DL framework testing methods }
     \label{fig:workflow}
\end{figure}

\textbf{Mutation-based Testing Methods.} These methods introduce mutation operators to modify the weight, parameter setting, and structure of models to generate mutants as the new test inputs. As shown in Fig.~\ref{fig:workflow}, the inputs of these methods are two pools consisting of different types of mutation operators and seed models or mutants, respectively.
They select one seed model from the model pool and one mutation operator from the operator pool, and then randomly select one middle layer of the model to mutate. 
After getting the generated mutants, they input the test data to the mutants and conduct differential testing on the result of the inferred calculation across different frameworks. Specifically, they focus on detecting three kinds of defects~\cite{wang2020lemon,gu2022muffin,guo2020audee,pham2019cradle}:
(1) Crash Defects. It manifests as the crash or exception appearing in the testing process and does not appear in all frameworks.  
(2) NAN Defects. It manifests as the output of the mutant containing NAN or INF and does not appear in all frameworks.
(3) Inconsistency Defects. It manifests the output inconsistency exceeding the preset threshold across different frameworks. 

The details of the state-of-the-art mutation-based testing methods~\cite{pham2019cradle,wang2020lemon,guo2020audee,li2023comet} are as follows.
CRADLE~\cite{pham2019cradle} is the earliest method to adopt public DL models and analyze their inference result inconsistency across different DL frameworks to detect defects. 
The subsequent methods LEMON~\cite{wang2020lemon} and AUDEE~\cite{guo2020audee} are two methods that mutate the original model to generate mutants for testing the framework. LEMON adopts structure-mutation and weight-mutation and combines MCMC~\cite{Andrieu2004AnIT} strategy to guide model mutation. AUDEE adopts input-mutation and weight-mutation and combines GA strategy to guide model mutation. COMET~\cite{li2023comet} is the latest method and covers all types of mutation operators used in LEMON and AUDEE, and combines three coverage metrics: layer sequence, layer input, and layer parameter to guide model mutation. 

\textbf{Template-based Testing Methods.} These methods focus on abstracting constraints or model structural templates to generate new models without mutating publicly available models and then detect framework defects across different DL frameworks. The key difference between these methods and mutation-based testing methods is whether using publicly available models. 
Gu et al.~\cite{gu2022muffin} proposed Muffin based on the directed acyclic graph theory to generate new models. It detects the defects during three processes of DL model execution: forward propagation, loss calculation, and reverse calculation. 
Liu et al.~\cite{liu2023generation} proposed Gandalf to generate diverse models by adopting context syntax constraints and combining the DQN strategy. Besides, Gandalf introduced 15 metamorphic relationships to enhance the framework testing effectiveness.
Liu et al.~\cite{liu2023neuri} proposed Neuri, which generates models via inductive rule inference. It first collects the invoking relationships of framework APIs to generate new models that can explore the deeper execution behavior of DL frameworks. Compared with mutation-based testing methods, template-based testing methods can cover more framework interfaces and generate more diverse model structures, but they also face the same challenges: many generated models contain random structures that are rare in real industry applications and cannot trigger defects that developers are concerned about.

\subsection{Mutation Testing}
\label{sec:mutationtesting}

This technology was initially proposed by Lipton et al.~\cite{mutate1971} in 1971 and further expanded by DeMillo et al.~\cite{DeMillo1978HintsOT} in 1978. It is applied early to evaluate the effectiveness of test inputs. Specifically, it injects common bugs made by programmers into the test program to generate defective versions of the test program and evaluates the quality of test input based on execution results. Recently, It has also been expanded to generate more new test inputs and widely adopted in several fields~\cite{Fioraldi2020AFLC,Padhye2018SemanticFW,Feng2021SnipuzzBF,Arora2016ASR,Habibi2015EventdrivenWA,Chen2016Coverage2,Andrieu2004AnIT}: Arora et al.~\cite{Arora2016ASR} generated test inputs through mutations to fully cover the execution paths of the concurrent programs, alleviating the challenges of low testing adequacy. 
Feng et al.~\cite{Feng2021SnipuzzBF} analyzed the input messages of IoT applications to design mutation operators and then mutated the existing input messages to generate new test inputs.
The Java class fuzz tool proposed by Chen et al.~\cite{Chen2016Coverage2} mutates candidate Java class files and combines them with MCTS algorithm\cite{Coulom2006EfficientSA} to guide the mutation process.  

Meanwhile, mutation testing has also been migrated to DL framework testing. Specifically,
researchers~\cite{pham2019cradle,wang2020lemon,guo2020audee,li2023comet} have introduced mutation technologies to generate more diverse DL models by changing the model structure, modifying the parameter setting, and editing neuron weights. In our work, we conclude the practical expertise of developers on DL model development to design new mutation operators and constraints for guiding mutation and filtering meaningless mutants. The fundamental difference between existing methods and our work is the motivation of the mutation operators: they aim to enhance test input diversity (e.g., structure diversity), whereas we focus on simulating the common operations of developers and detecting the defects close to real scenes.

\section{Methodology}
\label{sec:methodlogy}


\subsection{Overview}
\label{sec:overview}

\begin{figure}[]
     \centering
    \includegraphics[width=0.8\textwidth]{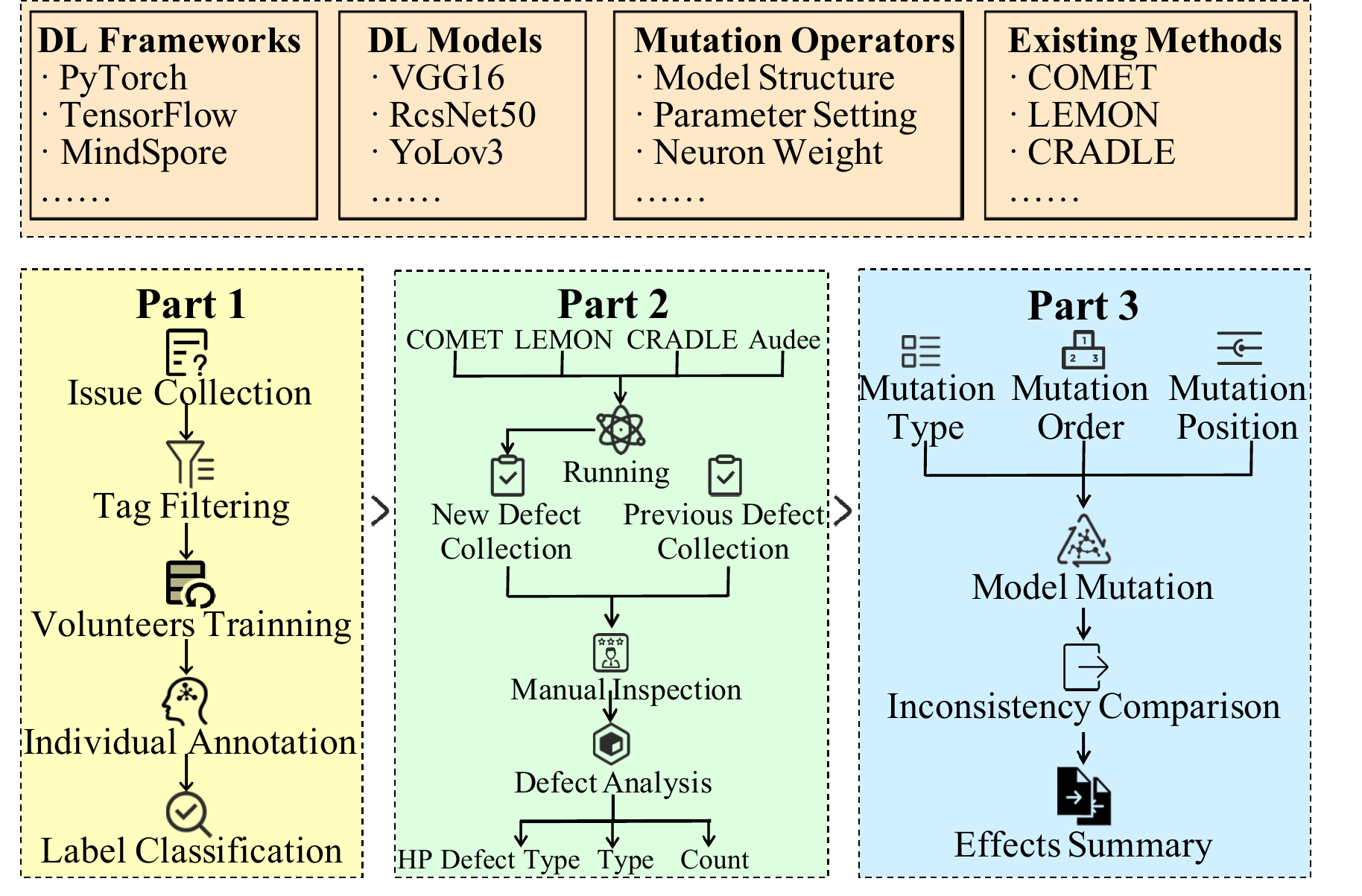}
     \caption{Overview of our methodology }    
     \label{fig:jiagou}
\end{figure}

This study aims to investigate the limitations of defect detection abilities of existing mutation-based testing methods and explore the relevant impact factors. 
Specifically, we investigate the defects considered important by the developers, evaluate the defect detection ability of existing methods, and further investigate the factors that affect the mutants since the mutants are key to existing methods for detecting framework defects.
To achieve the above goals, our study is divided into three parts as shown in Fig.~\ref{fig:jiagou}: 
(1) collect defect reports from DL framework communities and categorize them into different groups based on the expertise of developers;
(2) investigate defect reports exposed in running existing methods on new DL frameworks together with the previous defect reports based on the HP defect taxonomy constructed in the last part.
(3) analyze the output inconsistency of mutants generated under the different settings of the investigated factors.
Specifically, the first part constructs the evaluation criteria for the second part, which evaluates the defect detection abilities of existing methods. The third part further investigates the factors that affect the generated model and their specific effects, which can be used for designing our optimization strategies.




\subsection{High priority defect classification} 
\label{sec:hpdefectlabel}
\label{sec:3.1}

This section presents how to collect defect reports and label HP defects (i.e. Part 1 of Fig.~\ref{fig:jiagou}).
Specifically, we first crawl the issue reports from the framework communities and then select the HP defect reports based on specific tags. We recruit six volunteers to label the selected HP defect reports separately.

\begin{figure}[]
     \centering
     \footnotesize
    \includegraphics[width=0.8\textwidth]{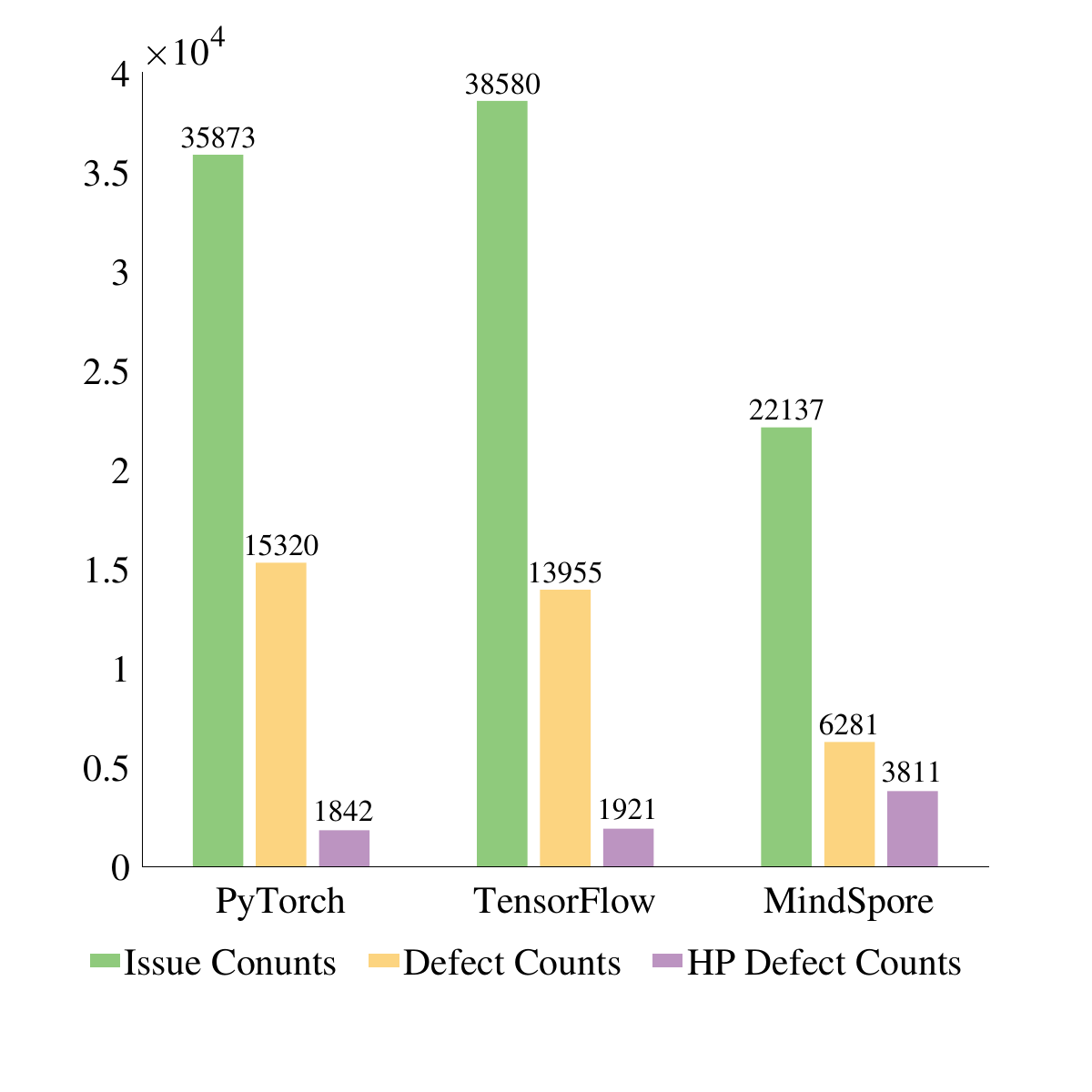}
     \vspace{-8mm}
     \caption{The statistics of collected issue reports }
     \label{fig:issueinfo}
\end{figure}

\textbf{Issue Collection and Tag Filtering.} We first collect issue reports from GitHub for ONNX~\cite{onnx}, TensorFlow~\cite{tfissue} and PyTorch~\cite{ptissue}, and Gitee for MindSpore~\cite{msissue}.
We notice that the developers often label the defect reports with specific tags (e.g., ``Serious'', ``type bug'', ``high priority'') to distinguish whether it is a defect and assign the priority for solving.
Therefore, we adopt the tags to exclude the issue reports irrelevant to framework defects, such as questioning, and filter the HP defect reports. Volunteers independently select defect tags from the three DL framework communities, then discuss and choose tags that reflect developers' high fix priorities. The selected tags are sent to developers for confirmation to ensure they can accurately represent priority.
Specifically, we use ``type bug'' as the filtering label for TensorFlow; use ``high priority'' and ``bug'' for PyTorch; use ``Main'', ``Serious'', and ``kind/bug'' for MindSpore. 
Notice that we manually check 1,500 defect reports tagged as ``type bug'' in TensorFlow based on symptom descriptions and developer comments, finding that over 90\% are considered high priority by developers. This proves the reliability of the collected defect reports.
The total counts of HP defects, all the defects, and all the issue reports are shown in Fig.~\ref{fig:issueinfo}.

\textbf{Volunteers Training.} To ensure the reliability of the labeling results and reduce the expensive label efforts 1,000 HP defect reports for each framework are sampled for volunteers to label.
Volunteers consist of five master's students and one Ph.D. student. They have engaged in testing DL frameworks like PyTorch, TensorFlow, and MindSpore and have been familiar with analyzing the framework defects for at least one year. 
To minimize misunderstandings and ensure accuracy, all volunteers are trained before labeling.
One volunteer first selects 200 defects from each framework to construct a preliminary HP defect classification and labels the defects based on the major type (i.e., the primary objects appearing in defect reports, such as functionality, performance, and resources) and the symptom description. 
This is because defect symptoms are more practical for classification as they are relevant to the testing process and often analyzed by testers. They provide observable information for quick and accurate identification, while root causes are rarely detailed in reports and require extensive developer analysis efforts.
Then all the volunteers hold an offline meeting to discuss the typical examples of each type and reach a consensus about the preliminary defect classification list. 
The typical examples help volunteers unify their understanding of each defect type in the preliminary list. These examples should (1) include keywords similar to the defect type in the symptom description and (2) contain explicit root cause analysis. The top 5\% of the most relevant examples, uniformly labeled by all volunteers, are then selected for discussion.



\textbf{Individual Annotation and Label Classification.} All the HP defects are randomly and evenly allocated to the volunteers. To minimize the potential subjectivity that affects the reliability of results, all the volunteers conduct a cross-validation labeling process for all the HP defects and discuss when necessary to resolve disagreements between each other.
They first identify the general type of defects based on the major types (e.g., functionality, performance, and resource). Then they determine the specific type of defect based on the symptom descriptions. For example, when the volunteers identify keywords such as ``CPU'', ``GPU'', and ``Memory'', they determine that the defect belongs to resource scheduling. Then, the volunteers further label the final type based on the symptom description like ``Core Dumped''.
When a volunteer encounters new types of defects not covered in the initial classification, he will inform others for discussion. The new type is added only when all the volunteers reach an agreement. All volunteers discuss the controversial or uncertain labeling results again, eliminate differences in these defects, and determine the final type. After achieving the labeling results, we calculate the $Kappa$ metric~\cite{fleiss1971measuring} to measure the labeling consistency of volunteers. The value reaches 0.804 on all labeled data, while it reaches 0.817, 0.777, and 0.815 on MindSpore, PyTorch, and TensorFlow data, respectively. All the values of the $Kappa$ metric exceed 0.7, proving the reliability of the labeling results~\cite{fleiss1971measuring}.
After completing the labeling processes, we conduct manual inspections to check the correctness. Among the labeled 3,000 defect reports, the accuracy of HP defect reports is 95\% after manual inspection. We collect the major type and brief symptom description of the HP defect types based on the labeling results after excluding the 5\% false positives as shown in Table\mbox{~\ref{tab:issue_category}}. We send the HP defect types to developers to check whether they are of high priority from their perspective. 

We ultimately obtained HP defect types for seven major types and 21 sub-types, referring to the installation, deployment, functionality, document, and execution of the framework. Specifically, the first column represents the summarized major types of HP defects, and the second column shows the specific symptom descriptions. Compared to the existing taxonomy, which includes three defect types (introduced in Section~\ref{sec:background}), our taxonomy classifies defects by more fine-grained symptom types (e.g., training crashes, compile crashes) and covers additional detection subjects such as resource scheduling and model performance, which the existing taxonomy ignores. 
Besides, defect types in our taxonomy are of high priority from the perspective of developers, this can provide more diverse references for researchers to understand framework defects. 

\begin{table}[]
  \centering
  \caption{HP defect types, corresponding descriptions, and their IDs}
  \resizebox{0.8\textwidth}{!}{
    \begin{tabular}{cll}
        \midrule    \textbf{Defect Type} & \textbf{Detailed Description} & \textbf{ID} \\
            \midrule
            \multirow{4}[1]{*}{Installation and deployment Defects} & PIP/Conda installation failure & A1 \\
                  & Docker installation/deployment failed & A2 \\
                  & Compatibility and dependency errors & A3 \\
                  & Lack of necessary environment support & A4 \\\midrule
            \multirow{3}[1]{*}{Resource scheduling Defect} & Memory/GPU resources error & B1 \\
                  & CPU underutilized & B2 \\
                  & Process/thread blocking or hanging & B3 \\
            \midrule
            \multirow{2}[2]{*}{Functional Defect} & Implementation errors or need optimize & C1 \\
                  & Lack functionalities or Version conflict & C2 \\
            \midrule
            \multirow{2}[2]{*}{ Wrong Description Defect} & Description error leads to failure & D1 \\
                  & Missing tutorial/explanation & D2 \\
            \midrule
            \multirow{3}[2]{*}{Performance Defect} & Slow execution,delay,timeout & E1 \\
                  & Abnormal loss value change & E2 \\
                  & Infer metric deteriorate & E3 \\
            \midrule
            \multirow{2}[2]{*}{Accuracy Defect} & Output inconsistency & F1 \\
                  & The returned result is incorrect or outlier & F2 \\
            \midrule
            \multirow{5}[2]{*}{Crash Defect} & Model construction, compilation failed & G1 \\
                  & Model Training/inferring errors & G2 \\
                  & Model Load/export error & G3 \\
                  & Interface Execution failed & G4 \\
                  & Complex framework program crash & G5 \\
            \midrule
            \end{tabular}
    }
  \label{tab:issue_category}
\end{table}

\subsection{Defect detection ability of existing methods }
\label{sec:defectdetection}


This section introduces how we evaluate the defect detection ability of existing methods (i.e., Part 2 of Fig.~\ref{fig:jiagou}). We first present how to collect the defects reported by existing methods and then show how we analyze them.

\textbf{Existing Methods Running and Defects Collection.} We run the existing methods on new DL frameworks with DL models and relevant test sets as introduced in Section~\ref{sec:benchmark} under the default settings as previous studies recommended, including the internal parameters of methods (e.g., the size of the generated model pool), test oracles with specific value settings of threshold, and the data processing strategies.
We collect the detected defects during the execution and the previous defects reported in their studies.

\textbf{ Defect Inspection.} We manually check the collected defects to exclude the false positives. Specifically, for the defects reported during the execution, we analyze the mutants that expose defects and eliminate those caused by illegal mutations. 
For the previous defects, we check their current status, i.e., whether some of them have been rejected till now. Then we further investigate the type and counts of the defects, especially the HP defects among them. 


\subsection{Investigation about the factors that affect mutants}
\label{sec:factorinvestigate}

This section presents how we investigate the specific effects of the factors on the mutants (i.e., Part 3 of Fig.~\ref{fig:jiagou}). 
We first introduce the details of the factors and then present the experiment about how to analyze the effects of these factors.

As shown in Fig.~\ref{fig:workflow}, each round of mutation starts from selecting mutation operators, one seed model, and one middle layer of the selected seed model. Then the mutation operator is adopted on the middle layer to conduct mutation. After finishing mutations, a new model is generated at each mutation round. By analyzing the mutation process, we can find the generation of mutants refers to three factors: mutation type (i.e. the type of mutation operators), mutation order (i.e., the order of mutation operators applied to the mutants), and the mutation position (i.e., the selected middle layer of the mutants to mutate). They all affect the generation of mutants and further affect the output inconsistency of the mutants across different frameworks, thus affecting defect detection. Therefore, we design the following comparative experiments to explore their specific effects.

\textbf{Factor Influence.} 
For the factor of mutation type, we conduct two experiments: (1) separately adopt each mutation operator for model mutation;
(2) adopt all mutation operators together as candidates for model mutation.
For the factor of mutation order, we analyze the mutants generated in different mutation orders based on the result of ``mutation type''; 
For the factor of mutation position, we divide the model structure into the ``backbone'' part and ``task head'' part based on the functionalities. The middle layers of the backbone are close to input data and extract data features while those of the task head are close to output data and execute specific tasks (e.g., image classification of face recognition) such as classification. We adopt each mutation operator separately on the middle layers of these two parts for model mutation. Notice that mutation rounds use different settings for each type of mutation operator due to their varied effects on models. Specifically, parameter mutation alters feature learning in the middle layers, weight mutation affects the output range of middle layers, input mutation changes the output scale, and structure mutation modifies the model's width and depth. Different mutation rounds are necessary to avoid generating empty models (e.g., LR deletes all layers) or invalid models (e.g., crashes or outputs beyond the effective accuracy range).

\textbf{Mutant Analysis and Effects Summary.} Finally, we analyze the output inconsistency of the mutants generated by different experiment settings and summarize how to control the factors for generating mutants that can effectively expose inconsistencies, thus contributing to defect detection, as shown in the last two steps of the third part in Fig.~\ref{fig:jiagou}.

\section{Experiment Design}
\label{sec:expsetup}

\subsection{Experimental Subjects}
\label{sec:benchmark}

\textbf{DL Frameworks.}
We collect the latest stable releases of 4 popular DL frameworks, i.e., PyTorch 1.13.1, MindSpore 2.2.0, TensorFlow 2.10.0, and ONNX 1.16.0 as the test objects. Among them, ONNX~\cite{onnx}, TensorFlow~\cite{tensorflow}, and PyTorch~\cite{torch} are most commonly adopted among existing methods~\cite{pham2019cradle,wang2020lemon,guo2020audee,li2023comet}, while MindSpore~\cite{mindspore} is a recently popular framework and widely welcomed by DL developers. 

\textbf{DL Models and Datasets.} To ensure the fairness and practical significance of our studies, we collect 23 DL models and corresponding test sets for eight types of tasks (e.g., image classification for face recognition and object detection for autonomous driving). The task type, model name, and the relevant test set name are shown in Table~\ref{tab:expinfor}.
Among them, 12 models and test sets are commonly adopted in existing methods~\cite{pham2019cradle,wang2020lemon,guo2020audee,li2023comet}, and the other 11 popular models and test sets marked with ``*'' are collected from the MindSpore and PyTorch communities and all widely used in industrial tasks such as autonomous driving~\cite{obermeyer2016predicting} and medical diagnosis~\cite{ChenSKX15}. We check these original models to ensure that their output differences under different frameworks are within 1e-4, which can ensure the correctness of our study.

\begin{table}[]
  \centering
  \caption{ Statistics of DL models and test sets}
    \resizebox{0.8\textwidth}{!} {
    \begin{tabular}{clc}
    \hline
    \textbf{Task Type} & \textbf{Model} & \textbf{Test Set} \\
    \hline
    \multirow{13}[2]{*}{Image Classification} & ResNet50-1* & CIFAR-10 \\
          & VGG16-1* & CIFAR-10 \\
          & AlexNet & CIFAR-10 \\
          & LeNet5-1 & Fashion-MNIST \\
          & LeNet5-2 & MNIST \\
          & ResNet50-2 & ImageNet \\
          & VGG19 & ImageNet \\
          & InceptionV3 & ImageNet \\
          & DenseNet121 & ImageNet \\
          & VGG16-2 & ImageNet \\
          & Xception & ImageNet \\
          & MobileNetV2 & ImageNet \\
\hline    \multirow{2}[2]{*}{Regression Prediction} & LSTM-1 & Sine-Wave \\
          & LSTM-2 & Stock-Price \\
    \hline
    \multirow{4}[2]{*}{Object Detection} 
          & SSD-resnet50-fpn* & COCO2017 \\
          & SSD-mobilenetv1* & COCO2017 \\
          & YoloV3* & COCO2014 \\
    \hline
    \multirow{2}[2]{*}{Semantic Segmentation} & DeeplabV3* & Pascal VOC \\
          & Unet*  & isbi\_challenge \\
    \hline
    NLP   & TextCNN* & Movie Review Data \\
    \hline
    Anomaly Detection & PatchCore* & MVTec AD \\
    \hline
    Defect Detection & SSIM-AE* & MVTec AD  \\
    \hline
    Key Point detection & OpenPose* & COCO 2017 \\
    \hline
    \end{tabular}
    }
  \label{tab:expinfor}%
\end{table}%

\subsection{Mutation-based testing methods}
\label{sec:baselines}
In our empirical study, we consider the following methods:

\begin{itemize}
    \item \textbf{CRADLE.} It is proposed by Pham et al.~\cite{pham2019cradle} and is the earliest work to detect framework defects by analyzing the DL models across different frameworks. 
    
    \item \textbf{AUDEE.} It is proposed by Guo et al.~\cite{guo2020audee} and adopts the genetic evolution strategy and mutates the inputs, parameters, and weights of the models to generate mutants. 
    
    \item \textbf{LEMON.} It is proposed by Wang et al.~\cite{wang2020lemon} and adopts the mutation operators in DeepMutation~\cite{ma2018deepmutation} and DeepMutation++~\cite{hu2019deepmutation++} and combines MCMC~\cite{Andrieu2004AnIT} strategy to guide the mutation. 

    \item \textbf{COMET.} It is proposed by Li et al.~\cite{li2023comet} and adds new mutation methods such as parameter mutation, tensor mutation, and outlier mutation based on LEMON. 
\end{itemize}

These methods are state-of-the-art mutation-based testing methods. To evaluate the defect detection abilities of these methods, we first gather defects detected in their previous studies. Then, we execute CRADLE, LEMON, and COMET on our benchmark to investigate whether they can detect new defects, as AUDEE is not open-source.

Moreover, we identify the main types of previous mutation operators, including input mutation, structural mutation, parameter mutation, and weight mutation, to investigate how they affect the model mutation of existing methods. 
The input-mutation includes changing the dimension and shape of layer input; 
the structural-mutation includes adding, deleting, exchanging, and replacing the layer; 
the parameter-mutation and weight-mutation include changing the parameters and weight values of the layer, as detailed in Table~\ref{tab:mutation}.

\begin{table}[]
  \large
  \centering
  \caption{ Mutation operators adopted in our study}
  \resizebox{0.95\textwidth}{!} {
    \begin{tabular}{ccc}
    \hline
    \textbf{Mutation Type} & \textbf{Mutation Operators} & \textbf{Description} \\
    \hline
    \multirow{3}[2]{*}{structure-mutation} & LA/LR & \multicolumn{1}{c}{Add/Delete one layer (inactive)} \\
          & LC/LS & \multicolumn{1}{c}{Copy one layer/Switch two layers (inactive)} \\
          & ARFm/ARFp & \multicolumn{1}{c}{Delete/replace an activation layer} \\
    \hline
    input-mutation & SM/DM & \multicolumn{1}{c}{Change the input shape/dimension of a layer} \\
    \hline
    parameter-mutation & PM    & \multicolumn{1}{c}{Changing the parameter value of a layer} \\
    \hline
    \multirow{3}[2]{*}{weight-mutation} & WS/NS & \multicolumn{1}{c}{Shuflle/Switch partial weights of a layer} \\
          & GF    & Add Gaussian noise to the one layer \\
          & NAI/NEB & Invert/eliminate partial weights of one layer \\
    \hline
    \end{tabular}%
    }
  \label{tab:mutation}%
\end{table}%


\subsection{Measurements and Test Oracles}
\label{sec:measures}


We measure the inconsistency degree by calculating the output distances of generated models across different DL frameworks since it is suitable for all DL models with different kinds of industry applications as recommended in previous work~\cite{gu2022muffin}. 
The details are as follows: Given a DL model with $n$ layers $ f= \langle L_1, L_2, \cdots, L_n \rangle$ and an input tensor $x$, the output of the $i$-th layer is recorded as $ f_{L_i} (x)$. The formula for calculating the distance $D^{M,N}_{f_{L_i}} (x)$ between DL framework $M$ and DL framework $N$ on layer $i$ of DL model $f$ is: 

\begin{equation}
    D^{M,N}_{f_{L_i}} (x) = mean (|M_{f_{L_i}} (x) - N_{f_{L_i}} (x) |)
    \label{equ:layerdis} 
\end{equation}

Then, to capture inconsistency defects that may be exposed in the middle layer of the model, we use the inconsistency change rate of two consecutive layers as follows:

\begin{equation}
    R^{M,N}_{f_{L_i}} (x) = | \frac{D^{M,N}_{f_{L_i}} (x) - D^{M,N}_{pre} (x)}{D^{M,N}_{pre} (x) + \epsilon} |
    \label{equ:inconsistency} 
\end{equation}

$D^{M,N}_{pre} (x)$ is the output of the direct precursor layer of the $i$ layer. The $\epsilon$ is set to 1e-7 to avoid the denominator being 0. Please note that we apply absolute calculations to capture both increases and decreases in middle-layer output inconsistencies.

\textbf{Test Oracle.} This study focuses on detecting crash defects, NAN defects, and inconsistency defects followed by the mutation-based testing methods selected. 
Specifically, we use the crash defect oracles and NAN defect oracles as recommended by previous studies~\cite{pham2019cradle, wang2020lemon,guo2020audee, li2023comet} as shown in the following content (i.e., the \textbf{Test Oracle 1} and \textbf{Test Oracle 2}). However, we adopt one new test oracle to detect inconsistency defects instead of following the default oracles of previous studies~\cite{pham2019cradle, wang2020lemon,guo2020audee, li2023comet} since there expand new DL models with diverse tasks like object detection in our study. 
Models collected from various tasks (marked with ``*'' in Table~\ref{tab:expinfor}) differ from those classification ones in existing work in terms of structural complexity, parameter scale, and task type (e.g., image classification). 
Therefore, the test oracle used in previous studies~\cite{pham2019cradle, wang2020lemon,guo2020audee, li2023comet} that adopts the metric $D\_MAD$~\cite{pham2019cradle} to measure inconsistency and the threshold for detecting defects is not suitable for these new collected models since it may cause large amounts of false positives. We adopt the \textbf{Test Oracle 3} for the models newly collected, as it can effectively detect inconsistency defects regardless of task type, structural complexity, or parameter scale in other studies~\cite{gu2022muffin}.



\textit{Test Oracle 1:} A NAN defect has been detected when some frameworks generate outliers such as ``NAN'' and ``INF'' while loading or executing models and other frameworks do not.

\textit{Test Oracle 2:} A crash defect is detected when some frameworks throw exceptions when loading or executing models and others do not.

\textit{Test Oracle 3:} For the second type of model, we use the $R^{M,N}_{f_{L_i}} (x)$ and when its value exceeds the preset threshold $t$, it is considered to detect a defect. Based on our experimental results, $t$ is set to 1e3 for low false positives.

We manually review and remove false positives from reported defects, submitting the remaining ones to developers for further analysis.

\subsection{Platform}

This part describes the execution parameter settings and environment information of our study. All existing methods~\cite{pham2019cradle,wang2020lemon,li2023comet} are downloaded from their open-source repositories and no changes are made to the source code. We control the model and the test set of each experimental run by modifying the configuration file, and the other parameters are set according to their recommendation and the default settings of the origin repositories. Our experiment is run on Ubuntu 20.04.4 LTS (GNU/Linux 5.4.0-152 generic x86\_64), with Intel (R) Xeon (R) Gold 6226R CPU @2.90GHz and CUDA V11.1.105 NVIDIA Geforce RTX 3090 and 24GB of memory size. 

\section{Result Analysis}
\label{sec:expresult}


\subsection{HP Defect Classification}
\label{sec:rq1result}
\begin{figure}[]
     \center
      \includegraphics[scale=0.7]{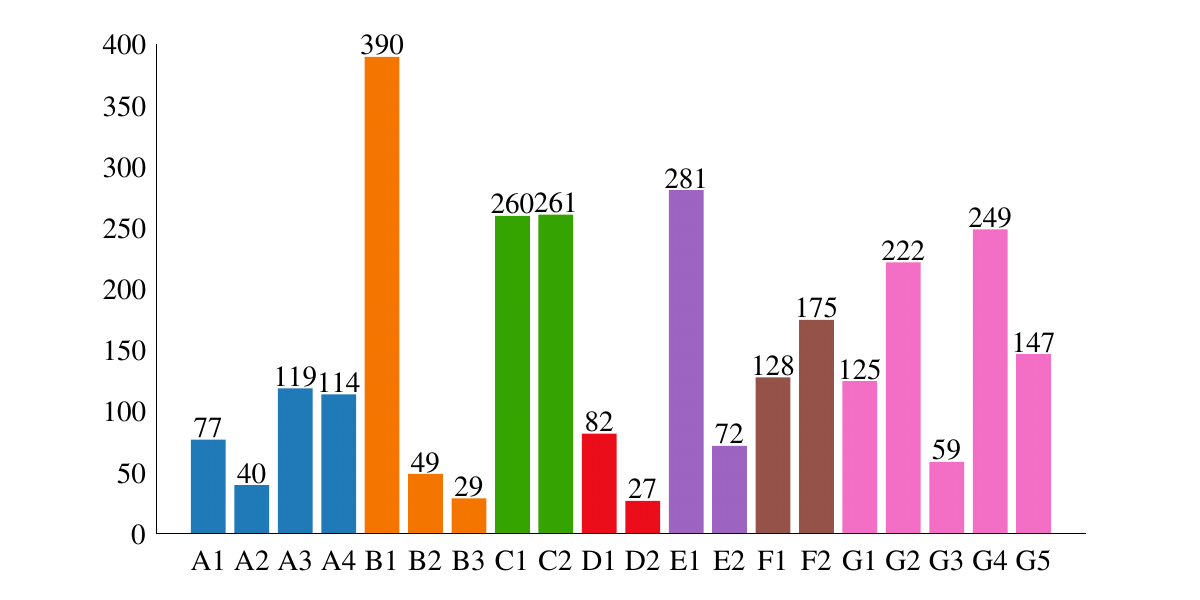}
     \caption{The distribution of labelled HP defect type}
     \label{fig:rq1dis}
\end{figure}

We first calculate the proportion of HP defects and investigate their general features. Fig.~\ref{fig:rq1dis} shows the counts of each type of HP defect with the $x$-axis representing the sub-types of each HP defect and the $y$-axis representing the defect count. The detailed analysis of HP defects is as follows.

\textbf{Installation and Deployment Defects (A)}. This type of defect includes four types of failure when installing or deploying frameworks. The failure of downloading (A1) and installing the framework using pip/conda and the failure of using framework docker (A2) account for 2.6\%. and 1.3\%, respectively. The defects related to the following failures: framework cannot be imported, conflicts with other Python tools (A3), and execution failures on specific operating systems and hardware devices (A4) account for 4.0\% and 3.8\%, respectively.

\textbf{Resource Scheduling Defects (B).} This type of defect mainly involves the wrong use of resources like memory, GPU, CPU, or other hardware resources. 
The defects related to the wrong use of memory, GPU, or other resources (B1) account for a total of 13\%. The specific symptoms of these defects are memory leaks, segmentation faults caused by the wrong allocation of memory or GPU resources, memory or GPU abnormal increases, untimely releases, and overflow or access errors.
The defects about low utilization, even idle, or unreasonable scheduling of CPU that lead to execution exceptions (B2) account for 0.97\%. 
The defects that cause blocking or hanging of framework programs (B3) account for 1.6\%. 

\finding{1}{
\textbf{Finding 1.} The most common sub-type of HP defects are the defects of managing resources like GPU and memory.
}

\textbf{Functional Defects (C)}. This type of defect mainly involves the mechanism level and interface level of the framework functionality. The interface defects mainly involve design and implementation errors and build failures on specific hardware devices. Besides, it also includes interface vulnerabilities that do not affect the normal execution but may be attacked (C1). These defects account for 8.67\%.
The function defects of the framework mechanism include incomplete implementation of existing functions, need for optimization, conflicts between the old version and the latest version, or lack of specific functions compared to other frameworks (C2), accounting for 8.7\%.

\textbf{Wrong Description Defect (D).} This type of defect mainly involves description errors in the documentation or missing necessary tutorials for users (D1), accounting for 3.63\%. The wrong description defect mainly includes cases where the users fail to build or run programs or obtain the wrong results due to wrong document descriptions, accounting for 2.73\%. The missing tutorial errors mainly include users requesting to provide necessary tutorial examples to solve common technical challenges that frequently occur (D2), accounting for 0.9\%.

\textbf{Performance Defects (E).} These defects include abnormal efficiency or evaluation metric values. The efficiency defects are related to the blocking, time out, or delaying during model training/inference or other framework programs or slow speeds compared to other frameworks (E1), accounting for 9.37\%. The evaluation metric defects are about the loss (E2), evaluation metrics like accuracy cannot reach the expected standards or are weaker than other frameworks (E3), accounting for 6.67\%.

\textbf{Accuracy Defects (F).} This type of defect mainly includes DL models and single interfaces.
The defects on DL models mainly involve the output, losses are abnormal (F1), accounting for 5.83\%. 
The defects on single interfaces include the interface output outliers such as NAN or the accuracy of the interface outputs that cannot meet the preset standard and expose a significant inconsistency in different operating environments or hardware devices (F2), accounting for 4.17\%.

\finding{1}{

\textbf{Finding 2.} Compared to accuracy defects, developers are more concerned about performance defects, such as loss and infer metric in model training and inferring.
}

\textbf{Crash Defects (G).}  This type of defect includes various crashes that cannot locate the root causes. The failure in build, compilation, save, and load of DL models (G1 and G3) accounts for 5.1\%. 7.4\% defects are related to unexpected crashes after running normally for a while in model training or inference (G2). The remaining 13.2\% defects are those complex framework program execution failures related to complex invocation scenarios (G4 and G5).

\finding{1}{
\textbf{Finding 3.} Crash defects are the most common major type of HP defects, especially the defects about complex invocations of framework interfaces.
}

\textbf{Summary.} From the perspective of the main types, crash defects account for the highest proportion, especially those related to the construction, deployment, and execution of DL models. 
From the perspective of sub-types, defects related to GPUs and memory are the HP defects with the highest proportion. 
This suggests that researchers should focus on the behavior of DL models in more diverse execution scenarios and pay attention to defects in execution processes such as resource scheduling. 
The A, C, and D types of HP defects indicate the user experience: whether the framework can be easily installed and deployed and whether the functions meet the expectations are concerned about by developers. 
Finally, both the accuracy and performance defects are related to the execution results, but the proportion of these two types of defects shows that developers are more concerned about the performance defects.





\subsection{Defect Detection Evaluation of Existing Methods}
\label{sec:rq2result}

This section reports the defect detection results of existing methods, especially the HP defects. 
The detection results for existing methods include (1) defects detected in previous studies and (2) defects detected during their execution of our new benchmark. Please note that the detected defects of AUDEE only include the previous ones since it is not open-source and cannot execute on our benchmark.
For the first kind of defects, we collect defects reported in previous studies and check their current status (i.e., whether the defect is confirmed or fixed) if there are open-source links available. 
For the second kind of defects, we run CRADLE, LEMON, and COMET for 100 rounds of mutation on all models followed by the settings of previous study~\cite{wang2020lemon} and analyze the defects reported by mutants generated in each round. To measure the output distance of generated mutants across different DL frameworks, we generate new mutants on MindSpore 2.2.0 and PyTorch 1.13.1 and adopt third-party tools TF2ONNX 1.16.0 and ONNX2Torch 1.5.13 to convert them to TensorFlow, and ONNX.
Please notice that the execution configuration and defect detection process follow the default settings of existing methods~\cite{pham2019cradle,wang2020lemon,li2023comet}. After excluding false positives, we further submit the defects to developers for confirmation. 
After gathering all collected defects, we analyze the total numbers of reported, confirmed, fixed, and HP defects identified by existing methods.


\begin{figure}[]
     \center
      \includegraphics[width=0.9\textwidth]{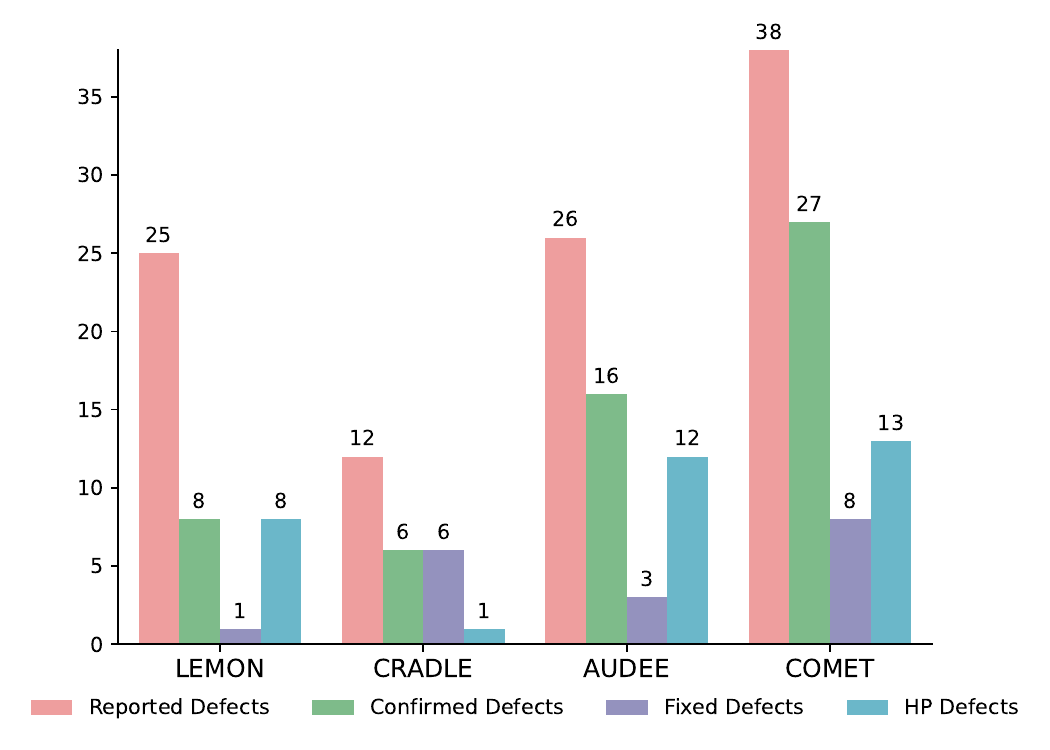}
     \caption{The number of defects detected by existing methods}
    \parbox{\textwidth}{\footnotesize * Please note that the defects counted in this figure consist of two parts: (1) defects detected in previous studies and (2) defects detected during the experiments on our benchmark.}
     \label{fig:rq2defect}
     
\end{figure}

\begin{figure}[]
     \center
      \includegraphics[width=0.9\textwidth]{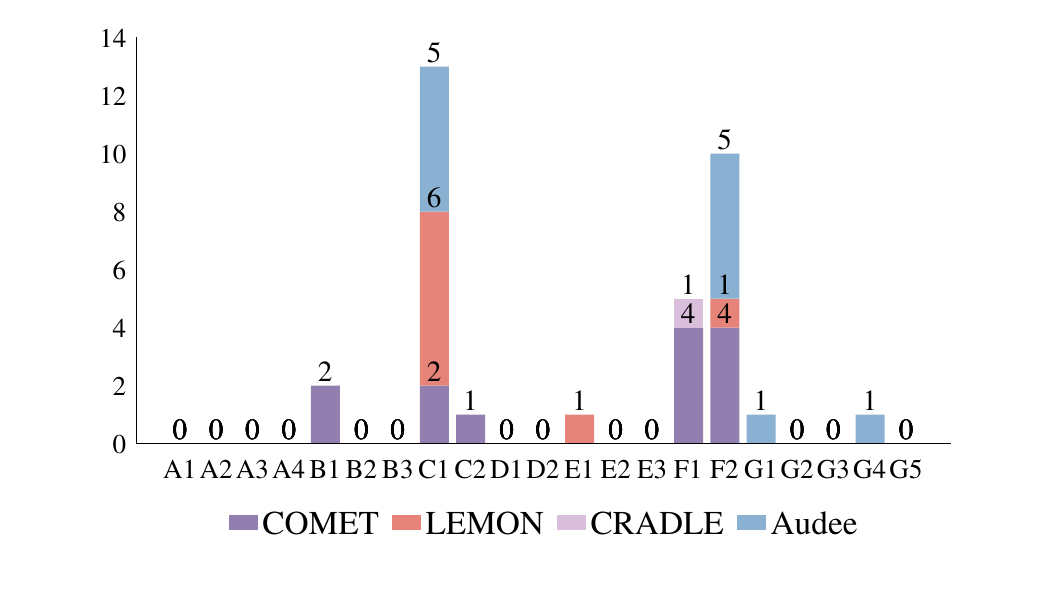}
     \caption{The number of HP defects detected by existing methods}
      \parbox{\textwidth}{\footnotesize * Please note that we only list the HP defects of each method in this figure. The total HP defects detected by one method in this figure correspond to the blue bar count for the same method in Fig.~\ref{fig:rq2defect}.}
     \label{fig:rq2hpdefect}
\end{figure}

\textbf{ General Defect Detection Results.}
Fig.~\ref{fig:rq2defect} shows the detected defect numbers of methods, with the $x$-axis representing the names of existing methods, and the $y$-axis representing the number of defects detected, confirmed, fixed, and HP defects. Please note that the blue bar in Fig.~\ref{fig:rq2defect} shows the total number of HP defects among the confirmed defects (i.e., the green bar in Fig.~\ref{fig:rq2defect}) after our analysis, and the details are in the next paragraph.
CRADLE does not report any defects during running on the new DL frameworks with new DL models, 
while it reports seven inconsistency defects across CNTK, Theano, and TensorFlow, and five model crashes in their studies. 
LEMON reports no inconsistency defects, two NAN defects, and two crash defects. Only one NAN defect has been confirmed by developers while another one is false positive. The two crash defects are false positives after the manual check. Besides, LEMON reports 13 inconsistency defects, one efficiency defect, six interface implementation defects, and four NAN defects in their studies, and only seven defects are confirmed by the developers. 
AUDEE detects 26 defects in their studies, of which 16 are confirmed, namely four inconsistency defects, five NAN defects, two model construction failure defects, and five interface implementation defects.  
COMET reports no inconsistency defects, seven NAN defects, and 320 crash defects in our study. Among them, seven NAN defects are defects in ``Conv2d'', ``pad'', ``mul'', and ``depthwise'' interfaces, so we submitted four defects, all of which are confirmed. After manual inspection, only six unique crash defects remain, of which three have been confirmed and one has been fixed. Besides, COMET reports 28 defects, of which 20 are confirmed by developers, including five inconsistency defects, three wrong output defects, one NAN defect, seven interface functional defects, two conversion failure defects, and two ``Core Dump'' defects.

\finding{1}{
 \textbf{Finding 4.} Existing methods rarely detect inconsistency defects, while reporting many duplicate crash defects and NAN defects.
}

As discussed in Section~\ref{sec:hpdefectlabel}, the crash defects, NAN defects, and inconsistency defects focused by existing work cover more coarse-grained symptoms while only part of such three kinds of defects belong to the HP defects. Therefore, we further investigate how many defects reported in the last part are HP ones based on the HP defect types identified in RQ1 to evaluate their ability to detect HP defects.

\textbf{HP Defect Detection Results.}
Fig.~\ref{fig:rq2hpdefect} shows the distribution of different kinds of HP defects detected by each method. The $x$-axis represents the sub-types of HP defects, and the $y$-axis represents the count of HP defects detected by existing methods. The confirmation process for HP defects detected by existing methods involves the following steps. We first analyze defects reported and confirmed by existing methods against the HP defect types in Table~\ref{tab:issue_category} to collect potential HP defects. Then, we consult with industry partners experienced in framework testing to evaluate these identified HP defects. Finally, we contact developers to confirm whether these HP defects are considered as of high priority. To date, developers have confirmed 13 defects are of high priority, with five already fixed.
Among the 34 detected defects, the two resource defects B1 involve Keras and ONNX interfaces throwing ``Core Dump'' exceptions. 13 interface defects C1 expose incorrect parameter settings in framework interfaces, leading to crashes, while one interface defect C2 related to missing functionalities in framework interfaces. The construction and execution defects, i.e., G1 and G4, cause the failure of the GRU model and ``ConvLSTM2D'' interface on the CNTK framework. The efficiency defect E1 involves repeated calls to the Keras ``clone'' function, causing slower execution. Five inconsistency defects F1 result from interface and model output inconsistency on different platforms (e.g., CPU and GPU). Finally, six outlier defects F2 involve NAN outputs from framework interfaces, and four defects produce incorrect outputs.

\finding{1}{
 \textbf{Finding 5.} Existing methods can detect HP defects in 5 major types and 8 sub-types, mainly including framework interface implementation defects and NAN defects.
}

We also analyze the missing HP defect types not reported by existing methods, i.e., those HP defects that they cannot detect. 
Firstly, existing methods cannot detect HP defects of types A (i.e. deployment defect) and D (i.e. wrong description defect) because they cannot be deployed in new environments, such as new hardware (e.g., Ascend), devices (e.g., autonomous driving systems), or operating systems (e.g., macOS). This limitation is due to their inability to perform model transformation and parameter adaptation in these environments. Additionally, existing methods focus on modifying seed model structures, overlooking the specific requirements of new environments (e.g., higher computing efficiency in autonomous driving systems), preventing them from triggering framework defects in these environments.
Secondly, existing methods cannot detect performance defects during training or inference processes or scheduling defects related to the CPU because they focus primarily on the correctness of execution results. Finally, existing methods can detect crash defects of G2, G3, and G5 types that may cause failures in loading weights or execution during model inference but cannot detect those in model training. 


\finding{1}{
 \textbf{Finding 6.} Among missing sub-types of HP defect, the detection ability of existing methods for performance defects related to model training is the weakest.
}

\subsection{Investigation about Mutation Influence Factors}
\label{sec:rq3result}

We conduct three types of experiments and use the metric $D^{M,N}_{f_{L_i}} (x)$ defined in Section~\ref{sec:measures} to measure output inconsistency across different frameworks. We investigate three factors: mutation type, order, and position. The key experimental parameter is the number of mutation rounds. Please note that the experimental settings of RQ2 cannot be directly migrated, as its experiments use a mixed selection of mutation operators guided by external strategies like MCMC~\cite{Andrieu2004AnIT} while we individually apply single mutation operators on seed models in the current RQ. we conduct experiments with different settings of mutation rounds since unsuitable mutation rounds may lead to the generation of invalid models. 
For example, if the mutation rounds exceed the depth of seed models, the ``LR'' operator will generate empty models since it will delete all the middle layers. Besides, the ``AFRm'' and ``AFRp'' operators cannot be applied too many times since activation layers are rare in the seed models (e.g., activation layers only account for 28.3\% of all the middle layers in the ``VGG16-1'' model). Furthermore, we find that the mutants generated by specific mutation operators after too many rounds (e.g., the ``DeeplabV3'' model after executing 50 rounds of the ``LA'' mutation) may produce too large outputs that exceed the effective accuracy range (e.g., 1e36 of the float32 in MindSPore) and become invalid models.
Therefore, we set the mutation rounds based on the large-scale experiment results to reduce the generation of invalid mutants. 

Specifically, weight-mutation operators have 100 mutation rounds across all models. Input-mutation, parameter-mutation, and most structure-mutation operators (except LR and AFRm) have 10, 40, or 100 rounds, depending on the seed model. LR has 5, 20, or 50 rounds, and AFRm has 5, 10, or 20 rounds, based on seed model depth. More details about the mutation rounds for different models can be found on our website~\cite{sharelink}.

\begin{figure*}[]	
    \centering    
    \begin{subfigure}[b]{0.32\textwidth}
        \includegraphics[width=\textwidth]{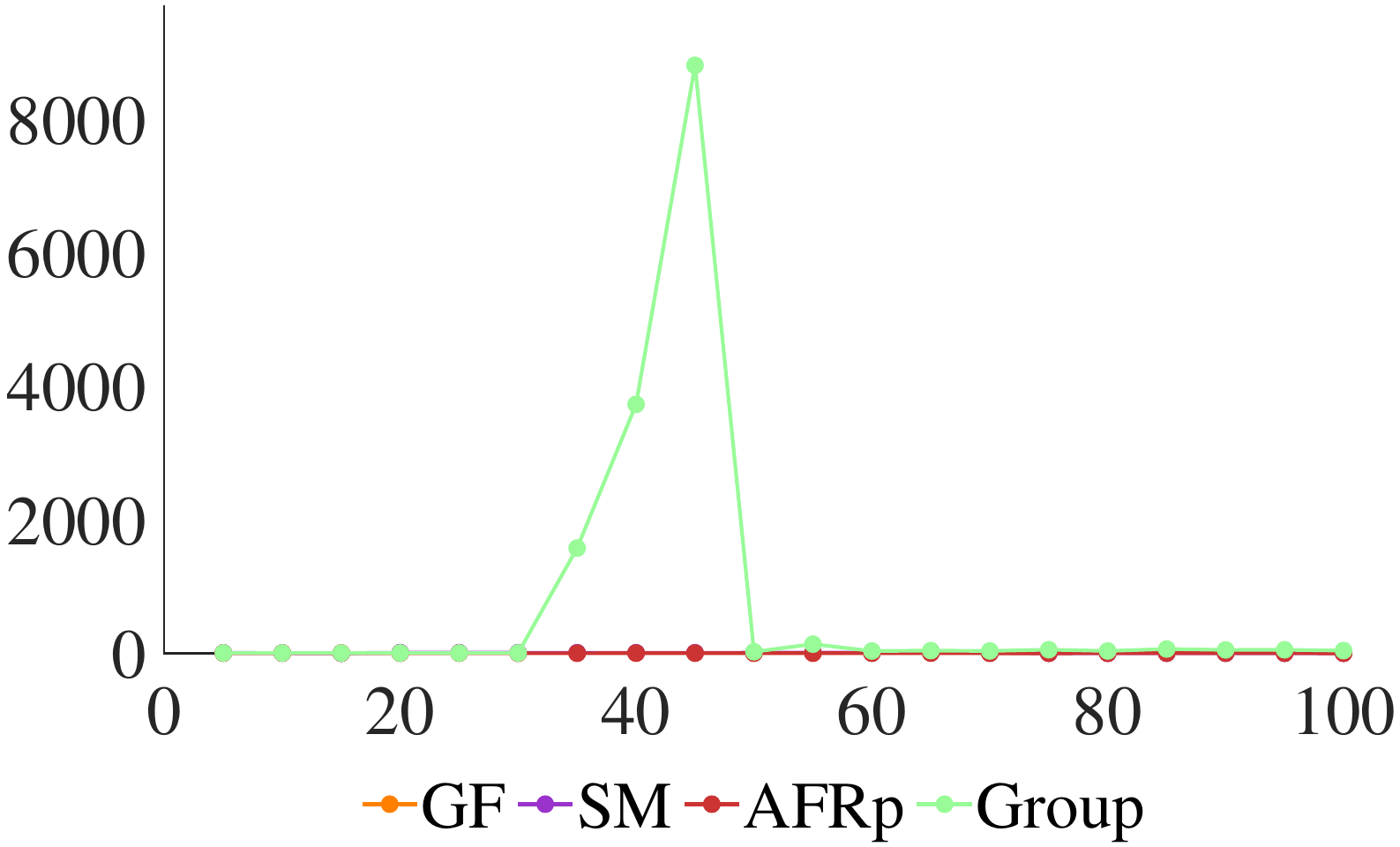}
        \caption{MindSpore and PyTorch}
        \label{fig:resnet500}
    \end{subfigure}
    \begin{subfigure}[b]{0.32\textwidth}
        \includegraphics[width=\textwidth]{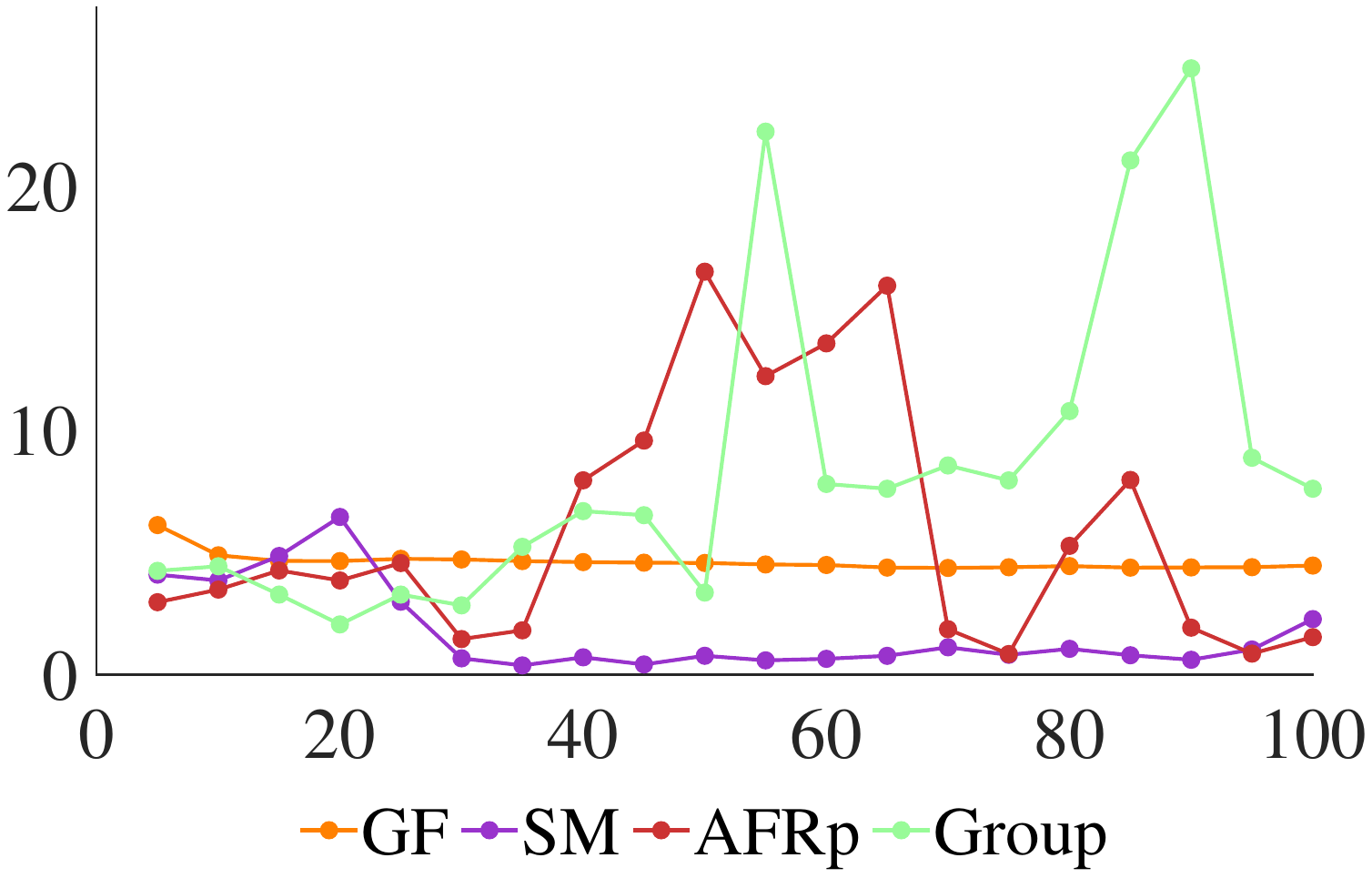}
        \caption{ONNX and PyTorch}
        \label{fig:resnet501}
    \end{subfigure}
    \begin{subfigure}[b]{0.32\textwidth}
        \includegraphics[width=\textwidth]{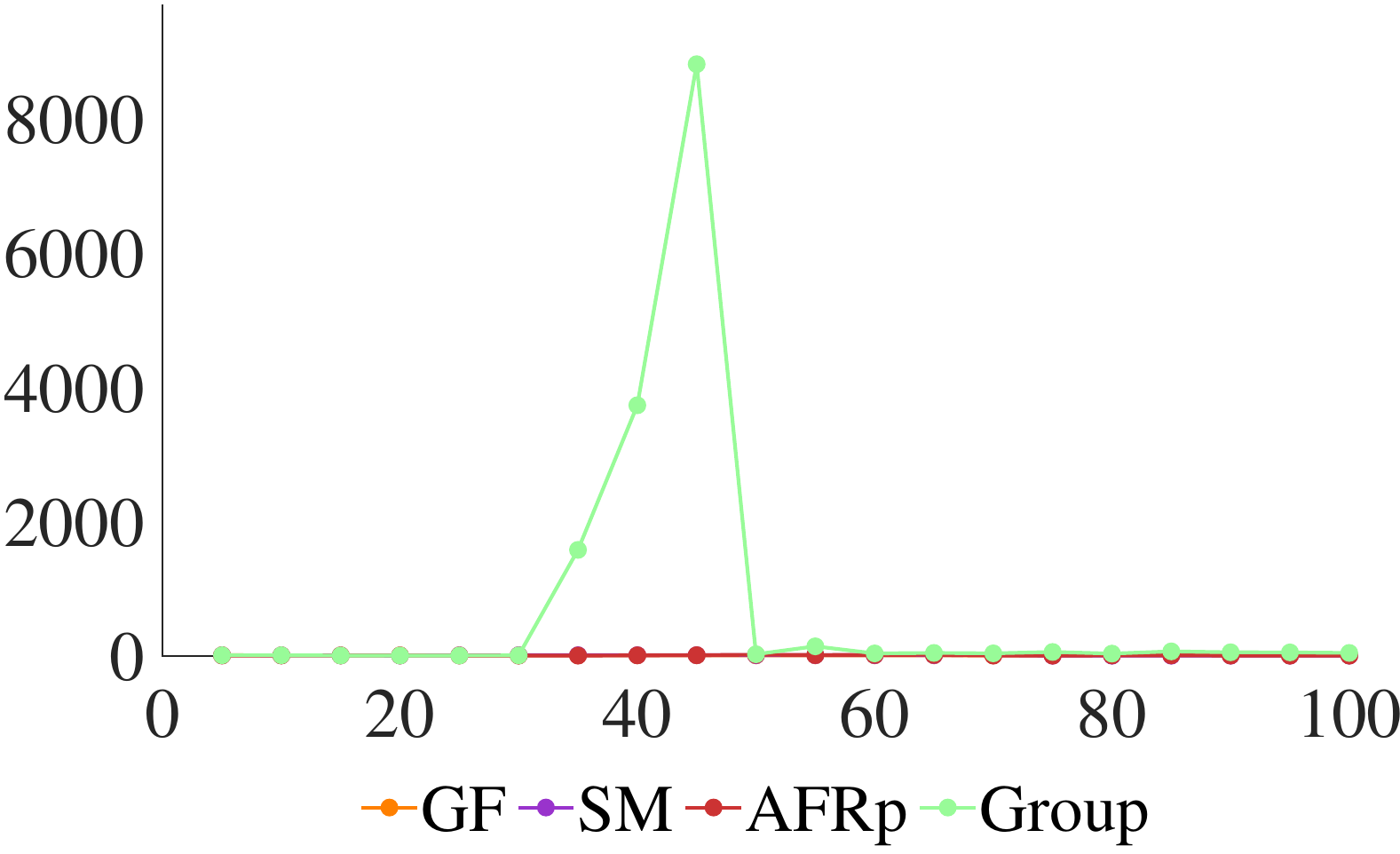}
        \caption{MindSpore and ONNX}
        \label{fig:resnet502}
    \end{subfigure}
	\caption{Output inconsistency of ResNet50-1 mutants generated by different types of mutation operators}
	\label{fig:rq3_1}
    \parbox{\textwidth}{\footnotesize * Please note that the lines in the above three subfigures represent the output inconsistency (it is calculated by the Formula~\ref{equ:inconsistency}) of mutants generated by specific mutation operators across two DL frameworks under different mutation rounds. For instance, in Fig.~\ref{fig:resnet501}, the red line highlights the output inconsistencies of ResNet50-1 mutants with the RA mutation operator under 100 mutation rounds across ONNX and PyTorch. We do not show the result of TensorFlow since the mutants of MindSpore, PyTorch, and ONNX all fail to convert to that of TensorFlow.}
\end{figure*}

\textbf{Result for Mutation Type.} 
We conclude the results about the proportion of illegal models, average execution time on all models, and the average output inconsistency of each mutation operator after excluding outliers as shown in the third to fifth column in Table~\ref{tab:rq3_1}. The values of each cell in the fifth column represent the average output inconsistency of the generated models across MindSpore and PyTorch, ONNX and PyTorch frameworks, and Mindpore and ONNX. Besides,
Fig.~\ref{fig:rq3_1} shows the most representative results of each type of mutation operator across three frameworks. The $x$ axis represents the mutation rounds, and the $y$ axis represents the output inconsistency of the generated models. 
The ``Group'' curve represents the mutants generated by multiple types of mutation operators. 
Regarding the average output inconsistency, the structure-mutation operators outperform other mutation operators (e.g., the red curve of the structure-mutation operator AFRp in Fig.~\ref{fig:resnet501}), while the weight-mutation operators perform worse (e.g., the orange curve of the structure-mutation operator GF in Fig.~\ref{fig:resnet501}). Besides, the structure-mutation operators can generate models that expose larger output inconsistency earlier (about 20 rounds) than others with an average output inconsistency as shown in ~\ref{fig:resnet501}. However, the parameter-mutation, i.e., the PM operator, tends to generate more false positives, i.e., illegal models, with a rate of 36.3\% higher than other mutation operators due to unreasonable changes in layer parameters.
Besides, input-mutation operators require more time than others with a maximum average time of 1836.59 seconds, while the average time of all other mutation operators is less than 1000 seconds. 
Using the ResNet50-1 model as an example, we observe that more input mutations increase the middle layer output size (54 rounds in SM and 57 in DM experiments). Overall, the total middle layer output size of the 100th mutant is 34.13\% larger than the original model. This introduces a larger scale of input data calculation, which needs more resources (e.g., GPU resources) and execution time.
Such operators can be applied to detect efficiency defects, i.e., the efficiency defect E1 in Table~\ref{tab:issue_category} since the mutants generated by the input-mutation operators refer to the calculation of large-scale tensor data and can obviously reflect the execution efficiency of the framework interfaces in DL models. 
The weight-mutation operators WS, NS, and NEB do not change the value range of the model weights (i.e, they do not adjust the maximum and minimum values of the weights) and can not generate mutants that can trigger larger output inconsistency than other kinds of mutation operators.
However, the GF mutation operator tends to generate models that produce outliers like NAN since it can enlarge the value of model weights with the increase of mutation rounds, which can be applied to detect outlier defect F2 in Table~\ref{tab:issue_category}.

\begin{table}[]
  \centering
  \caption{Statistic results about mutation type}
  \large
  \resizebox{\linewidth}{!}{
    \begin{tabular}{ccccc}
    \toprule
    \textbf{Mutation Type} & \textbf{Mutation Operator} & \textbf{Illegal Rate} & \textbf{Execution Time (s)} & \textbf{Average Output Inconsistency} \\
    \midrule
    \multirow{6}[2]{*}{structure-mutation} & LA    & 24.20\% & 586.818  & (409.489,195.694,336.967) \\
          & LR    & 15.90\% & 264.818  & (0.296,0.814,0.826) \\
          & LC    & 17.50\% & 583.909  & (68.411,20.000,104.313) \\
          & LS    & 15.90\% & 360.273  & (2.556,16.317,16.175) \\
          & AFRm  & 17.80\% & 164.200  & (215.331,167.590,333.939) \\
          & ARFp  & 15.90\% & 464.182  & (0.658,1.982,1.935) \\
    \midrule
    \multirow{2}[2]{*}{input-mutation} & SM    & 15.90\% & 1540.636  & (0.795,0.784,1.117) \\
          & DM    & 26.70\% & 2132.545  & (0.467,0.985,0.890) \\
    \midrule
    parameter-mutation & PM    & 36.30\% & 550.750  & (37.303,352.990,22.780) \\
    \midrule
    \multirow{5}[2]{*}{weight-mutation} & WS    & 18.20\% & 642.091  & (0.527,0.748,1.010) \\
          & NS    & 18.20\% & 362.000  & (0.842,2.124,1.997) \\
          & GF    & 30.90\% & 944.545  & (0.994,2.751,2.725) \\
          & NAI   & 18.20\% & 358.818  & (5.833,5.352,6.541) \\
          & NEB   & 18.20\% & 399.455  & (0.899,1.670,2.080) \\
    \midrule
    multiple kinds of mutations & Group & 27.20\% & 994.100  & (118.634,70.401,153.320) \\

    \bottomrule
    \end{tabular}%
    }
  \label{tab:rq3_1}%
\end{table}%

\finding{1}{
 \textbf{Finding 7.} The structure-mutation operators outperform others in model mutation.
 
 \textbf{Finding 8.} The input-mutation operators can detect defects related to efficiency and resource allocation.
}

		


\textbf{Results for Mutation Order.} 
Except for the result of the single kind of mutation in the last part,
we also carry out 100 rounds of mutation with multiple mutation operators as candidates on seed model.
We analyze the output inconsistency between high-order or low-order mutation of the models generated by the single mutation and multiple kinds of mutations.

Compared with different kinds of a single mutation, multiple kinds of mutations can reach higher peaks of output inconsistency as shown in Fig.~\ref{fig:resnet501}. However, it eventually decreases or even reaches zero caused by adopting mutation operators that reduce the model structure complexity like the LR operator as shown in Fig.~\ref{fig:resnet500} and Fig.~\ref{fig:resnet502}.
Meanwhile, the effect of using multiple mutation operators may not be better than using a single mutation operator during early mutation stages. However, as the mutation order increases, using multiple mutation operators is more likely to produce mutants that can effectively expose output inconsistency compared to a single mutation operator, as shown in Fig.~\ref{fig:resnet501}. 
This shows multiple mutation types can generate mutants with larger output inconsistency faster than single mutation operators in most models. 
Furthermore, high-order mutations are more likely to produce mutants that can lead to false positives. For instance, 66.6\% of the results show that all NAN mutants are generated in the last 50\% of mutations, and they are all false positives after manual inspection.


\finding{1}{
 \textbf{Finding 9.} Low-order mutation outperforms high-order mutation and considering higher mutation orders tend to introduce more false positives.
}

\begin{figure*}[]	
    \centering    
    \begin{subfigure}[b]{0.33\textwidth}
        \includegraphics[width=\textwidth]{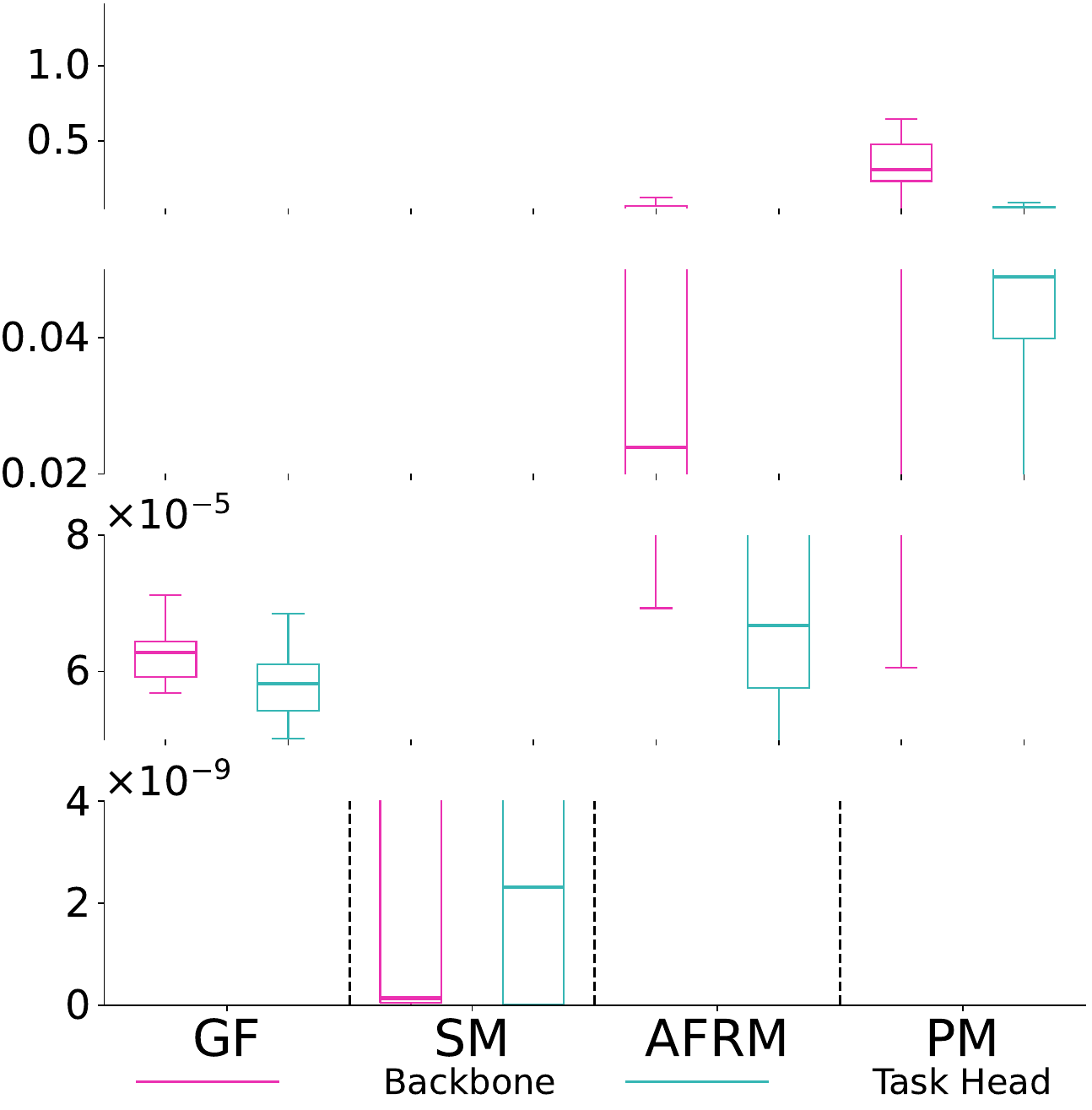}
        \caption{MindSpore and PyTorch}
        \label{fig:ssimae0}
    \end{subfigure}
    \begin{subfigure}[b]{0.32\textwidth}
        \includegraphics[width=\textwidth]{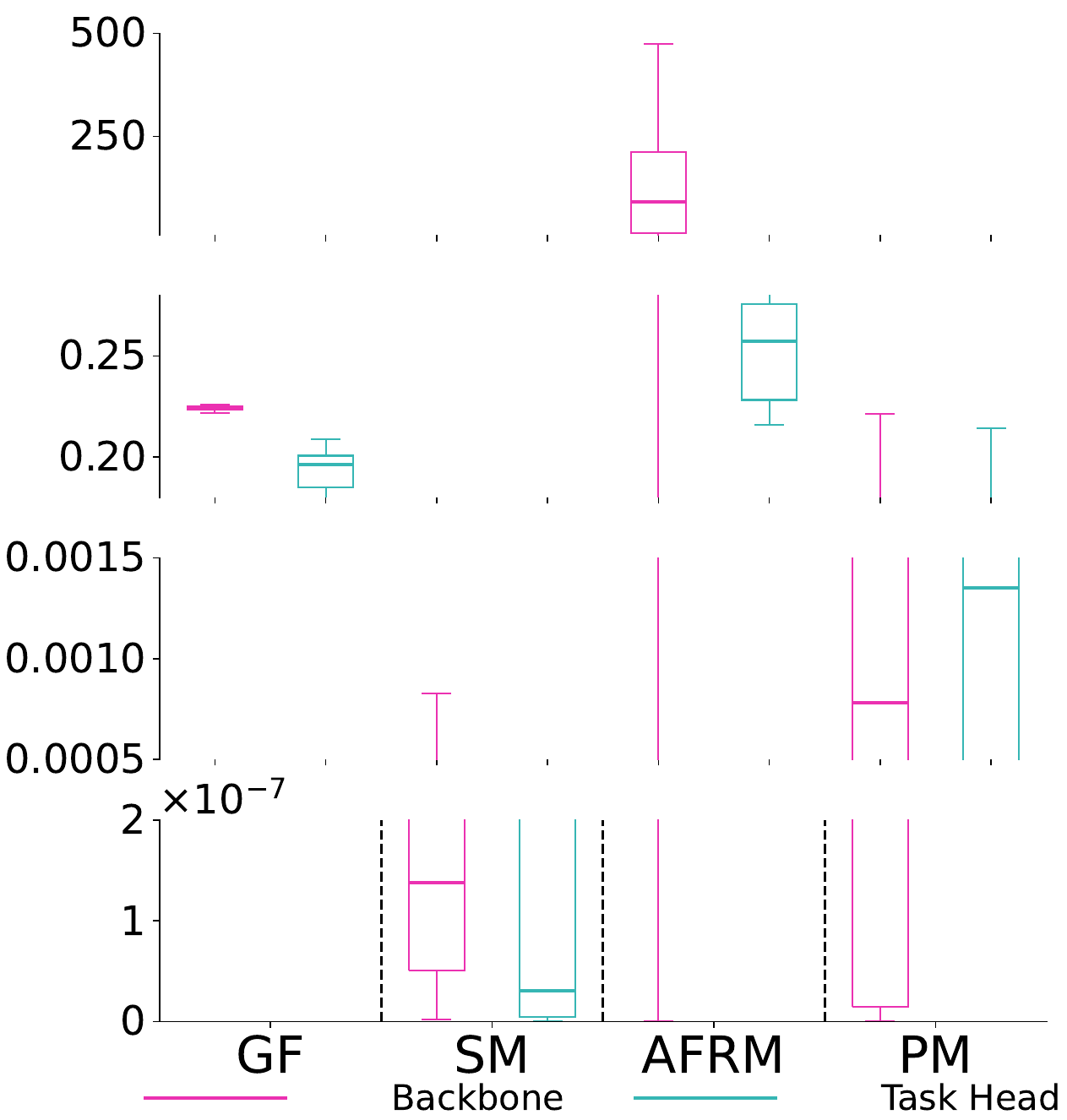}
        \caption{ONNX and PyTorch}
        \label{fig:ssimae1}
    \end{subfigure}
    \begin{subfigure}[b]{0.33\textwidth}
        \includegraphics[width=\textwidth]{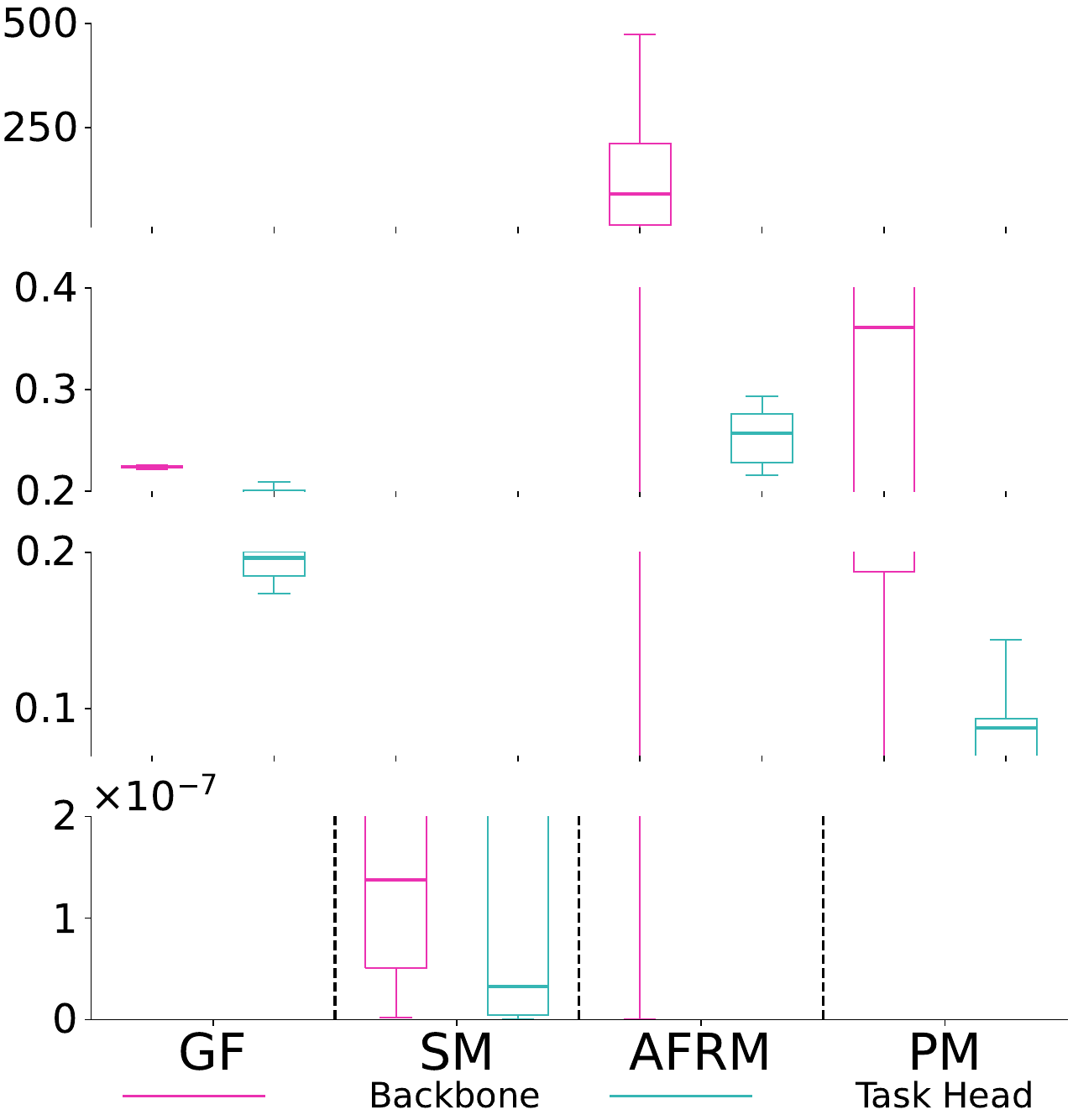}
        \caption{MindSpore and ONNX}
        \label{fig:ssimae2}
    \end{subfigure}
	\caption{Comparison results about the output inconsistency of SSIM-AE mutants generated in different mutation positions}

    \parbox{\textwidth}{\footnotesize * Please note that the boxes in the above three subfigures represent the output inconsistency (it is calculated by the Formula~\ref{equ:inconsistency}) of mutants generated by specific mutation operators across two DL frameworks. For instance, in Fig.~\ref{fig:ssimae0}, the red box highlights the output inconsistencies of SSIM-AE mutants with mutations (GF, SM, AFRM, and PM) in the backbone's middle layers across MindSpore and PyTorch, while the blue box indicates inconsistencies for mutants with mutations (GF, SM, AFRM, and PM) in the task head's middle layers. We do not show the result of TensorFlow since the mutants of MindSpore, PyTorch, and ONNX all fail to convert to that of TensorFlow.}

	\label{fig:rq3_3}

\end{figure*}

\textbf{Results for Mutation Position.} 
We separately adopt 14 mutation operators on the middle layers of the backbone and task head and compare the output inconsistency of generated models, respectively. Due to the clearest backbone and task head range division of the ``VGG16-1'', ``OpenPose'', ``DeeplabV3'', and ``SSIM-AE'', we conduct experiments on these models and show the result of ``SSIM-AE'' in Fig.~\ref{fig:rq3_3} due to the paper limitation. The $x$ axis represents the best one in different kinds of mutation operators, and the $y$ axis represents the output inconsistency of the models generated by mutating in the backbone or task head. We divide the box in Fig.~\ref{fig:rq3_3} into three parts with different value ranges due to the significant output inconsistencies across models generated in different mutation orders. As shown in Fig.~\ref{fig:rq3_3}, the average output inconsistency of ``backbone'' is higher than that of ``task head'' in ten groups and only loses in two groups. Specifically, mutating in the backbone shows a higher rate of 61.54\% in total results than in the task head. This reveals mutating in the backbone part can trigger greater output inconsistency than those in the task head parts. Besides, we can also find the distribution range of the backbone is larger than that of the task head, which means that mutating in the backbone can generate more diverse models with varying degrees of exposure inconsistency.


\finding{1}{
\textbf{Finding 10.} Mutating the layers of the backbone of models outperforms mutating in the task head of models.
}

\section{Discussion}
\label{sec:discussion}

\subsection{Optimization Strategies on Existing Methods} 
As shown in Fig.~\ref{fig:workflow}, the executing process of existing mutation-based testing methods can be divided into two parts: (1)  model mutation and (2) defect detection. We propose optimization strategies for these two parts separately.

\textbf{Model Mutation.} Humbatova et al.~\cite{humbatova2021deepcrime} simulate real faults in DL models using mutation operators. Motivated by this, we propose new mutation operators to simulate common developer operations and enhance constraints to reduce illegal mutants combined with Findings 7-10. This strategy aims to explore more valuable test input spaces and detect defects closer to real industry scenarios. We present two specific optimization strategies. Researchers first need to summarize the common operations of users when developing DL models based on frameworks, and then abstract them into specific mutation operators.
Besides, researchers need to analyze the development expertise of developers that help to identify illegal models and design new constraints to avoid generating and filtering them. They must also design new constraints to select mutation operators and positions and control the mutation orders.

\textbf{Defect Detection.} Based on Findings 1-6, we propose to expand the scope both for the detection process and detection subjects. The process includes more stages in the DL model lifecycle, including model construction, execution, and deployment. The subjects refer to state parameters during the model lifecycle, like resource usage, execution time, and loss values.
This strategy aims to test more scenarios for the framework and enrich the types of framework defects detected. Specifically, researchers should 
focus on more subjects (e.g., resource
usage) and defect types that are frequently exposed in real scenes and have severe impacts on framework users during model lifecycles to design corresponding test oracles and detect diverse defects.

\begin{table}[htbp]
  \centering
  \caption{Details about Defects Detected after Optimization}
  \resizebox{\textwidth}{!}{
    \begin{tabular}{ccccc}
    \toprule
    Id & Description & Confirmed & Fixed & Type \\
    \midrule
    1     & The ``Relu6'' interface cannot correctly deal with ``NAN'' values. & $\checkmark$  & $\times$ & \multirow{4}[2]{*}{F2} \\
    2     & Relu6 operator processing nan value logic error & $\checkmark$  & $\times$ &  \\
    3     & nn.Batchnorm2d interface outputs with nan & $\checkmark$  & $\times$ &  \\
    4     & Defect about the output of Conv2d interface is nan & $\checkmark$  & $\checkmark$  &  \\
    \midrule
    5     & Defects in the ``Add'' interface and ``Mul'' operators of MindSpore & $\checkmark$  & $\times$ & F1 \\
    \midrule
    6     & nn.pad execution failed & $\checkmark$  & $\checkmark$  & E1 \\
    \midrule
    7     & nn.Flatten operator memory allocation failure defect & $\checkmark$  & $\checkmark$  & B1 \\
    \bottomrule
    \end{tabular}%
    }
  \label{tab:defectopt}%
\end{table}%

\subsection{Defect Detection after Optimization}
Based on the optimization findings, we modify the settings of COMET and execute the new version on the benchmarks since COMET is the latest mutation-based testing method now. Finally, we detect seven new defects of four types, including four NAN defects (that is, the F2 type in Table~\ref{tab:issue_category}), one resource scheduling defect (i.e., the B1 type in Table~\ref{tab:issue_category}), one efficiency defect (i.e., the E1 type in Table~\ref{tab:issue_category}), and one inconsistency defect (i.e., the F1 type in Table~\ref{tab:issue_category}). All of them are confirmed by developers and three are fixed. Details about these defects are shown in Table~\ref{tab:defectopt} and can also be found on our website~\cite{sharelink}. Next, we present three typical defect cases to show how we detect new defects.

\textbf{Case 1.} We conduct 50 mutations on VGG16 using SM and DM mutation operators to amplify data shape and dimension. During execution, the generated mutant of MindSpore encounters a memory allocation failure, while the equivalent PyTorch model can run normally. This is because the DM mutation expands the input data dimension in the backbone's final layer of VGG16, causing execution errors in the ``Pad'' interface (it is inserted to ensure the correctness of the input data shape of the middle layers during mutation). Developers have confirmed that it is caused by the implementation error of the MindSpore Pad'' operator on the secure memory setting.

\textbf{Case 2.} We select the GF operator and perform 100 mutations on the structure ``ResNet'' which is the backbone part of the model PatchCore. In the 6th generation mutation, we find that the output of the mutants has already changed to NAN for both MindSpore and PyTorch. However, when the running environment is changed from GPU v100 to GPU A100, the output changes from ``NAN'' to ``inf''. During the reproduction process, developers discover that when running the mutant on the Ascend 910, the output becomes a normal number, and the underlying framework mechanism's output design is incorrect.

\textbf{Case 3.} We select SM and perform 50 mutations on the ``Openpose'' and find that the calculation process of the MindSpore mutant after the execution is significantly slower than that of the PyTorch. After preliminary analysis, the calculation time is mainly spent on the ``Flatten'' operator introduced by SM. After confirmation by the developers, it is because the ``Flatten'' operator of MindSpore executes 45.86\% slower than that of PyTorch. The reason is that the implementation of the ``Flatten'' operator in MindSpore lacks the necessary optimization, and the current version does not support this ability. It is confirmed by developers that it will be fixed in future versions.


\section{Threats to Validity}
\label{sec:threats}

\textbf{Internal Threats to Validity.} This kind of threat comes from the correctness of our manual labeling and the implementation of existing mutation-based methods. To reduce the bias introduced by the subjectivity of volunteers, six volunteers with at least one year of development experience individually label the defect reports. Besides, we adopt the $Kappa$ metric~\cite{fleiss1971measuring} metric to measure the consistency of the label results while the average value on three DL frameworks reaches 0.803, representing the high consistency among them. We further send the typical cases of each kind of HP defect to developers for further confirmation and to achieve a positive response.
To investigate the real performance of existing methods, we collect the public version of baseline methods and execute them under default parameter settings without any modification.

\textbf{External Threats to Validity.} This kind of threat comes from the collected defect reports from different DL frameworks and the benchmark we used, including the DL frameworks to be tested and the DL models with relevant test data. To address this, we systematically collect data from three DL frameworks, covering all defect reports from their initial release to the present, ensuring comprehensive analysis.
Besides, we adopt three popular DL frameworks as targets and collect 23 DL models that have been widely used in previous studies (12 models) or applied in real industry applications (11 models). We plan to extend more diverse DL frameworks (e.g., Jittor~\cite{jittor}) and introduce more new kinds of DL models with different tasks (e.g., speech recognition) to validate the generality of our findings.

\textbf{Construct Threats to Validity.}  This kind of threat mainly comes from the reliability of selected HP defect reports.
To alleviate this threat, we adopt the specific defect tags to conduct filtering since the crawled issue reports contain noise data like questions, demands, and other reports that are not defect reports. All volunteers independently select defect tags from the three DL framework communities, after which they discuss and agree on the tags that reflect the developers' high fix priority. The selected tags are then confirmed by developers to ensure they accurately represent the priority. Besides, the volunteers further conduct manual inspections to check whether the selected defect reports are of high priority based on the predetermined list.



\section{Related Work}
\label{sec:relatedwork}

\subsection{API-Level DL Framework Testing Methods}

DL frameworks encapsulate a large number of complex functionalities into operators for users. Therefore, researchers also conduct studies on testing these framework interfaces. Based on the kinds of how to detect defects, existing methods can be divided into (1) differential testing methods, (2) metamorphic testing methods, (3) and LLM-based testing methods.

\textbf{Differential Testing Methods.} 
This kind of method generates new test inputs by adjusting parameter settings or data, then analyzes execution results across modes (e.g., CPU and GPU) to detect bugs. It also aligns different framework interfaces with specific parameter settings or reimplements them to ensure consistent outputs, integrating them into new test inputs for detecting bugs.
Deng et al.~\cite{Deng2022FuzzingDL} test those framework interfaces with high input/output parameter similarity by automatically inferring their relationships based on the semantics and then analyzing the consistency of the execution state of interfaces. 
Yang et al.~\cite{Yang2023FuzzingAD} focus on testing those interfaces related to automatic differential calculation. They abstract such interfaces as the functions for processing tensors and then compare the accuracy of their gradient calculation results to detect bugs. 
Wei et al.~\cite{Wei2022FreeLF} propose Freefuzz, which mines the input of the framework interface from technical communities and wild code fragments. Then, it mutates the collected original input and tests the target framework interface by checking the consistency between the CPU execution result and that of the GPU. 
Zhang et al.~\cite{zhang2021predoo} propose Predoo, a DL operator accuracy testing method that mutates the test input of DL operators to extend the numerical accuracy to maximize the detection of accuracy errors in DL frameworks across CPU and GPU.
Duo~\cite{zhang2021duo} optimizes test input generation and testing with two mutation methods and strategic algorithms. It combines differential and fuzz testing, using nine mutation operators to evaluate DL framework interfaces on TensorFlow, PyTorch, MNN, and MXNet for detecting bugs.

\textbf{Metamorphic Testing Methods.}  
These methods~\cite{wang2022eagle,ding2017validating,chen2024miss} analyze individual framework interfaces to design metamorphic relations (MRs) that constrain generated test inputs, focusing on parameter settings and inputs to detect bugs within a single framework.
Wang et al.~\cite{wang2022eagle} proposed EAGLE, which generates equivalent test inputs based on 16 equivalent rules for framework interfaces, and then detects bugs by analyzing the consistency of the execution results between equivalent test inputs. 
Chen et al.~\cite{chen2024miss} proposed Meta, a framework interface testing method, which can efficiently detect interface implementation errors and accuracy bugs guided with 18 MRs designed for modifying interface parameters and input tensors.
Ding et al.~\cite{ding2017validating} proposed 11 MRs related to data transformation from the train/validation/test set. The MRs require that the classification accuracy of DL models remain consistent after training or inference using the modified data.

\textbf{LLM-Based Testing Methods.} With the rapid development of large language models (LLM), researchers also have attempted to apply LLMs~\cite{deng2023large1, deng2023large2} for generating new test inputs. Deng et al.~\cite{deng2023large1} propose the first LLM-based testing method, TitanFuzz. It produces the test cases and then mutates them based on the generative-style and infill-style prompts to LLMs, respectively. Lately, they find that TitanFuzz tends to generate test cases that are already existing and cannot explore newer state space. To solve this, Deng et al.~\cite{deng2023large2} subsequently proposed FuzzGPT, which adopts prompt or fine-tuning to make LLMs learn from historical test inputs automatically.

Compared with the model-level testing methods, the above API-level testing methods focus on testing single framework interfaces, including the data, parameter setting of interfaces. They can thoroughly analyze the execution behaviours of single interfaces while ignore the combinations of multiple interfaces, thus failing to detect defects exposed in the interactions between interfaces.

\subsection{Empirical Studies On DL Framework Bugs}

Framework users often report issues to communities, including the defect reports and technical questions. Some researchers~\cite{Chen2022TowardUD,jia2021symptoms,Makkouk2022AnES} mine the data from different communities to conclude findings for optimize DL frameworks. 

Some researchers summarize the symptoms, root causes, and repair patterns of framework defects to provide suggestions for investigating the defects~\cite{Chen2022TowardUD,jia2021symptoms,Makkouk2022AnES}.
They often mine the framework community's defect reports and technical blogs for their study.
Chen et al.~\cite{Chen2022TowardUD} identify 800 typical defects from 4 popular DL frameworks. Then they analyze their common features and provide important references for DL framework defect repair. 
Li et al.~\cite{jia2021symptoms} collect defect reports from TensorFlow and analyze their distribution in real DL applications. Then they give guidance for the detection and localization of common framework defects. 
Tarek Makkouk et al.~\cite{Makkouk2022AnES} focus on performance defects closely correlated to immediate feedback tasks and explore factors that affect the execution cost. 

Besides, researchers also reproduce the triggering conditions of defect reports from communities to analyze their internal causes and propagation processes, thus designing specific detection methods for different kinds of defects. 
Tambon et al.~\cite{tambon2024silent} conducts empirical research on ``silent'' defects on Keras and TensorFlow, which are difficult for detection and causes more concealed damage since it displaies no error messages and may not lead to crashing or hanging. 
Makkouk et al.~\cite{makkouk2022empirical} focus on performance defects closely related to real-time feedback tasks (e.g., autonomous driving) in DL frameworks. They collect performance defect reports of PyTorch and TensorFlow and then explore the factors that affect the computational execution cost.
Zhang et al.~\cite{zhang2020empirical} focus on the defects that cause DL applications (e.g., medical diagnosis systems) to crash after long term execution, analyze their distribution and root causes, and provide directions for repairing such defects.

Prior work aims to provide suggestions for detecting and repairing defects by investigating the symptoms, root causes, and characteristics of framework defects. In contrast, we focus on evaluating the defect detection ability of mutation-based testing methods and factors affecting their performance by analyzing the defects detected by existing mutation-based testing methods. Besides,
we further propose a new defect taxonomy from the perspective of the developers' fixing priority on defects and then provide feasible optimization suggestions, which can detect seven defects of four new types that existing methods cannot detect. 


\section{Conclusion}
\label{sec:conclusion}
This paper studies the status of DL frameworks.
Using this benchmark, we built a more practical benchmark based on developers' ratings and identified the limitations of existing methods.
Through our analysis, we propose further optimizations, which can be achieved by:
(1) design new mutation operators that can simulate the real practice of users;
(2) enhance the mutation constraints for generating more mutants that are more common in real scenes;
(3) detect more types of defects by analyzing more tasks such as model training and focusing on more subjects like efficiency, performance of models, and GPU/CPU management.

\section*{acknowlegement}
This work is supported partially by the National Natural Science Foundation of China (62372228), the Fundamental Research Funds for the Central Universities (14380029), and the Open Project of State Key Laboratory for Novel Software Technology at Nanjing University under (Grant No. KFKT2024B21).

\bibliographystyle{ACM-Reference-Format}
\bibliography{sample-base}

\end{document}